\documentclass[prd,aps,preprint,superscriptaddress,showkeys,showpacs,
 preprintnumbers,amsmath,amssymb,floatfix]{revtex4-1}
\usepackage{graphicx}
\usepackage{epsfig}
\usepackage{amsfonts}
\usepackage{dsfont}
\usepackage{amsmath}
\usepackage{amssymb}
\usepackage[sort&compress]{natbib}
\usepackage{dcolumn}
\usepackage{multirow}
\usepackage{bigstrut}
\usepackage{mathrsfs}
\usepackage{type1cm}
\usepackage{placeins}
\usepackage{color}
\usepackage{rotating}
\usepackage{slashed}
\usepackage{subfigure}
\usepackage[mathscr]{euscript}
\newcommand{\Adj}{\mathrm{Adj}}

\begin{document}

\title{Covariant Quantization of CPT-violating Photons}

\author{D. Colladay}
\author{P. McDonald}
\affiliation{New College of Florida, Sarasota, FL, 34243}

\author{J. P. Noordmans}
\author{R. Potting}
\affiliation{CENTRA, Departamento de F\'isica, Universidade do Algarve, 8005-139 Faro, Portugal}

\date{\today}
\vspace{3em}

\begin{abstract}
\noindent 
We perform the covariant canonical quantization of the CPT- and Lorentz-symmetry-violating photon sector of the minimal Standard-Model Extension,
which contains a general (timelike, lightlike, or spacelike) fixed background tensor $k_{AF}^\mu$. Well-known stability issues, arising from complex-valued energy states, are solved by introducing a small photon mass, orders of magnitude below current experimental bounds. We explicitly construct a covariant basis of polarization vectors, in which the photon field can be expanded. We proceed to derive the Feynman propagator and show that the theory is microcausal. Despite the occurrence of negative energies and vacuum-Cherenkov radiation, we do not find any runaway stability issues, because the energy remains bounded from below. An important observation is that the ordering of the roots of the dispersion relations is the same in any observer frame, which allows for a frame-independent condition that selects the correct branch of the dispersion relation. This turns out to be critical for the consistency of the quantization. To our knowledge, this is the first system for which quantization has consistently been performed, in spite of the fact that the theory contains negative energies in some observer frames.
\end{abstract}


\maketitle

\section{Introduction}

The covariance of physical laws under boosts and rotations is at the basis of the standard model (SM)
of particle physics and general relativity \cite{relativity}. This Lorentz symmetry is closely related
to the invariance under the combined action of charge conjugation, parity inversion, and time reversal, i.e. CPT symmetry \cite{cpt,Gre02}.
Over the last few decades the interest in the possibility of breaking Lorentz and CPT symmetry has been growing. This rise is
motivated by theories that attempt to unify general relativity with quantum mechanics and that exhibit mechanisms of 
Lorentz and CPT breaking~\cite{qgmodels1,qgmodels2,reviews}. The detection of a corresponding experimental signal would provide profound new physical insights and could point us to the correct theory of quantum gravity.

In this context, the Standard-Model Extension (SME) has proven to be a tool of great value. It is a framework that incorporates
Lorentz- and CPT-violating effects into the SM \cite{kostelecky-colladay}, gravity \cite{gravity}, and for matter-gravity couplings \cite{mattergravity}, by extending the Lagrangian to include all possible Lorentz- and CPT-violating operators consisting of the conventional fields. Because of its generality it allows for broad experimental searches \cite{datatables} as well as general theoretical considerations of Lorentz- and CPT-violating effects.

Of particular interest is the pure-gauge matter sector of the minimal SME, which includes only superficially renormalizable operators. Here, Lorentz violation (LV) can be introduced,
either while preserving CPT, or while violating it. In this paper we consider the Chern-Simons-like operator of mass dimension three that causes both CPT- and Lorentz-symmetry breaking, parametrized by
a fixed background vector $k_{AF}^\mu$ \cite{carroll-field-jackiw}. Although bounded observationally to minute levels
\cite{datatables, carroll-field-jackiw}, this term has received intensive attention in the literature, as it
arises as a radiative correction from the fermion sector in the
presence of a LV axial-vector term \cite{jackiw-kostelecky}. It is thus important
to establish both in the fermion and the photon sector, whether such effects impede a rigorous quantization, and if not,
in what way the standard procedures have to be modified. While the quantization of the fermion sector was implemented successfully in the past \cite{kostelecky-lehnert}, the situation is more ambiguous in the photon sector.
Furthermore, while CPT-violating effects are strongly bounded in the photon sector,
this is not at all the case in the gluon and weak gauge-boson sector \cite{datatables}.
Although we only consider the abelian case here, our analysis may lead to important implications for LV non-abelian theories as well.

In this paper, we thus perform the covariant quantization of Maxwell-Chern-Simons theory. Covariant quantization is extremely useful in performing
quantum-field-theoretic calculations, as the formulas retain explicit
covariance throughout the computational procedure.
In a previous work, it was discussed how this can be implemented
for the CPT-preserving case \cite{don-pat-potting1, hohensee}. In Ref.~\cite{don-pat-potting2} the quantization for purely timelike $k_{AF}^0$ was discussed and applied to calculate vacuum-Cherenkov-radiation rates, whereas in Ref.~\cite{adam-klinkhamer,andrianov},
attention was restricted to the (massless) case of purely spacelike $k_{AF}^\mu$ in an axial gauge. Although some of the present results were already presented in Ref.~\cite{don-pat-potting2}, the approach we take here is more general and rigorous, while we consider spacelike, lightlike, as well as timelike values of $k_{AF}^\mu$ in a general class of covariant gauges. 

As in Ref.~\cite{don-pat-potting1}, the introduction of a mass regulator turns out to be necessary for a consistent quantization. 
This is phenomenologically feasible, due to the fact that ultra-tight observational bounds on $k_{AF}^\mu$ \cite{datatables} allow the choice of a photon mass sufficiently large to fix quantization problems, while simultaneously agreeing with current experimental bounds.
 Furthermore, the introduction of a photon mass is often used 
to regulate infrared divergences that turn up in loop diagrams in both conventional calculations,
and in the context of LV effects \cite{cambiaso-lehnert-potting2}.
We note that the introduction of a photon mass in the context of the SME
has been studied in the presence of both CPT-preserving and
CPT-violating terms at the level of the equations of motion and
the propagator in Ref.~\cite{cambiaso-lehnert-potting}.

The outline of this paper is as follows. In section~\ref{Lagrintroduction} we introduce the Lorentz- and CPT-violating model, including a nonzero photon mass that is introduced through the St\"uckelberg mechanism. Subsequently, we
find a covariant basis of normalized and orthogonal eigenvectors of the equation-of-motion operator in section~\ref{sec:polvec}.
These polarization vectors satisfy modified orthogonality relations when fixed to be on shell.
In section~\ref{sec:disprel} we analyze the equations of motions
in momentum space,
and show that for the case of timelike $k_{AF}^\mu$ the introduction
of a nonzero mass parameter avoids a region in three-momentum space that has no corresonding real energy solutions. We also find a condition on $k^\mu_{AF}$ that guarantees energy positivity. Energy positivity and its connection to stability is further discussed in section~\ref{sec:energypositivity}, where we find a way to distinguish different branches of the dispersion relation in any observer frame. We derive a relation between the momentum-space propagator and the polarization vectors in section~\ref{sec:propandpol}.
The field operator is then quantized in terms of creation and annihilation
operators in section~\ref{quantization}. Subsequently, the commutator of fields at spacelike separation is worked out in section~\ref{causality},
and it is shown that the theory satisfies microcausality.
In section~\ref{feynprop} the Feynman propagator is derived and in section~\ref{brst} we analyze the space of states in the context of
BRST quantization.
Finally we present our conclusions in section~\ref{sec:conclusions}. Some of the more detailed analyses of the dispersion relation, the photon group velocity, and the energy lower bound are relegated to the appendices.

\section{CPT-violating Photon sector of the SME}\label{Lagrintroduction}%

CPT violation in the photon sector of the power-counting renormalizable part of the SME is given by the Lagrangian
\begin{equation}
\mathcal{L}_{A,k_{AF}} = -\frac14 F_{\mu\nu} F^{\mu\nu} +
\frac12 k_{AF}^\kappa \epsilon_{\kappa\lambda\mu\nu}A^\lambda F^{\mu\nu}\,,
\label{L-massless}
\end{equation}
where $k_{AF}^\mu$ is an arbitrary real-valued and fixed background vector with the dimensions of mass. The CPT-violating term in Eq.~\eqref{L-massless} is gauge invariant up to total derivative terms, which, in the absence of topological obstructions, do not influence the physics.

The theory in Eq.~\eqref{L-massless} breaks so-called particle Lorentz symmetry, while it is invariant under observer Lorentz transformations \cite{kostelecky-colladay}. Observer Lorentz transformations are just transformations of the coordinates of the reference frame of the observer and thus transform both $k_{AF}^\mu$ and the fields. Particle Lorentz transformations, on the other hand, affect only the particle fields, but leave the tensor $k_{AF}^\mu$ unchanged. This corresponds to changing the orientation and/or velocity of the experimental system in absolute space.

In this paper, we will consider the cases of spacelike, lightlike, and timelike $k_{AF}^\mu$.
As is well known \cite{carroll-field-jackiw},
for timelike $k_{AF}^\mu$ the dispersion relation following from
(\ref{L-massless}) has a tachyonic character: there are (small) momenta for which there are no
corresponding real solutions for the energy,
signaling an unstable theory that does not permit a consistent quantization (although proposals have been made for fermionic theories \cite{tachyonicfermions}).

As was noted first in \cite{alfaro-cambiaso},
a way around this problem is to introduce a small mass term for the photon
through the St\"uckelberg mechanism \cite{stueckelberg}.
The gauge-fixed photon Lagrangian becomes
\begin{equation}
\mathcal{L}_A = -\frac14 F_{\mu\nu} F^{\mu\nu} +
\frac12 k_{AF}^\kappa \epsilon_{\kappa\lambda\mu\nu}A^\lambda F^{\mu\nu}  
+ \frac12 m_\gamma^2 A_\mu A^\mu - \frac1{2\xi}(\partial_\mu A^\mu)^2\ ,
\label{lagrangian-A}
\end{equation}
where $\xi>0$ is a gauge-fixing parameter and $m_\gamma$ is the small photon mass. 
For spacelike, lightlike, as well as timelike values of $k_{AF}^\mu$, a nonzero photon mass is useful. In addition to its use as a regulator for infrared divergences \cite{cambiaso-lehnert-potting2}, it allows for the introduction of so-called concordant frames \cite{kostelecky-colladay}. 

Concordant frames are observer frames in which the LV effects can be treated as small perturbations to the Lorentz-symmetric physics. For nonzero values of $k_{AF}^\mu$, this cannot be the case in all frames, because $k_{AF}^\mu$ changes under the action observer Lorentz transformations. Therefore, the size of its components is in principle unbounded, if one allows for arbitrary observer frames. To be compatible with experimental constraints, Earth's restframe is then presumed to be in a concordant frame. However, to say anything meaningful about the size of the components of $k_{AF}^\mu$, we need a nonzero photon mass, since it is the only other dimensionful parameter in Eq.~\eqref{lagrangian-A}. As mentioned in the introduction, the required size of photon mass lies many orders of magnitude below its experimental bounds. We will discuss this in more detail in Section~\ref{sec:disprel}.

\section{Polarization vectors}\label{sec:polvec}
\label{eigenvectors}

In momentum space, the classical equation of motion, corresponding to the Lagrangian in Eq.~\eqref{lagrangian-A}, reads
\begin{equation}
\left[(p^2-m_\gamma^2)\eta^{\mu}_{\;\;\nu}-(1-\xi^{-1})p^\mu p_\nu
-2i\epsilon^{\alpha\beta\mu}_{\;\;\;\;\;\;\;\nu}(k_{AF})_\alpha p_\beta\right]
e^{(\lambda)\nu}(\vec p) \equiv S^{\mu}_{\;\;\nu} e^{(\lambda)\nu}(\vec p) = 0\ ,
\label{eom}
\end{equation}
where $e^{(\lambda)\nu}(\vec p)$ are the eigenvectors of the equation-of-motion operator $S^{\mu}_{\;\;\nu}$. The index $\lambda$ runs over $0,3,+,-$, labeling the gauge mode, and three physical modes, respectively. Contraction of Eq.~\eqref{eom} with $p_\mu$ yields 
\begin{equation}
(\xi^{-1} p^2 - m_\gamma^2) (p \cdot e^{(\lambda)}) = 0,
\end{equation}
demonstrating that there is a gauge mode satisfying $p^2 - \xi m_\gamma^2 = 0$.
This expression also establishes that the physical polarization vectors, corresponding to the remaining modes, satisfy $p\cdot e^{(\lambda)} = 0$.
The contraction of Eq.~\eqref{eom} with $(k_{AF})_\mu$ gives a similar expression that demonstrates
the fact that the physical polarization modes either obey the conventional dispersion relation $p^2 = m_\gamma^2$, or the corresponding polarization vectors satisfy
$k_{AF} \cdot e^{(\lambda)}=0$. These facts are confirmed by the explicit expressions for the polarization vectors in Eq.~\eqref{polarvect} and by the functions defining the dispersion relations in Eq.~\eqref{eigenvalues}.

When the eigenvectors $e^{(\lambda)\nu}(\vec p)$ satisfy Eq.~\eqref{eom}, they are functions of the three-momentum $\vec{p}$, since $p^0 = p^0(\vec{p})$ is fixed by the dispersion relation. As shown in Refs.~\cite{don-pat-potting1, don-pat-potting2}, quantization can be carried out referring only to these on-shell polarization vectors. However, it turns out to be useful to consider $e^{(\lambda)\nu}(p)$ as functions of both $p^0$ and $\vec{p}$ that satisfy
\begin{equation}
S^{\mu}_{\;\;\nu}e^{(\lambda)\nu}(p) = \Lambda_\lambda(p) e^{(\lambda)\mu}(p)\ ,
\label{eomofshell}
\end{equation}
where $\Lambda_\lambda(p)$ is the eigenvalue of $S^\mu_{\;\;\nu}$ belonging to the polarization mode $\lambda$. The relation
\begin{equation}
\Lambda_\lambda(p) = 0,
\end{equation}
can then be imposed to enforce the equation of motion. Each of the resulting dispersion relations has two solutions, corresponding to the conventional positive and negative energies. Usually, one then uses the positive root of $\Lambda_\lambda(p)$ to define the energy of the on-shell polarization vectors. However, in Section~\ref{sec:disprel} we show that in the present LV case, the sign of the roots of $\Lambda_+(p)$ is invariant only in concordant frames \cite{kostelecky-lehnert}, i.e. frames where the components of $k_{AF}^\mu$ are small compared to the photon mass. In non-concordant frames this sign can depend on the size and direction of $\vec{p}$. We will discuss this issue in more detail in Section~\ref{sec:disprel}. For now, we let $E_\lambda(\vec{p})$ denote the root of $\Lambda_\lambda(p)$ that is positive in a concordant frame. Substituting the solution $p^0 = E_\lambda(\vec{p})$ in the expression for $e^{(\lambda)\nu}(p)$ then gives the relevant on-shell polarization vector that satisfies the equation of motion in Eq.~\eqref{eom}.

We determine the explicit solutions for the polarization vectors $e^{(\lambda)\mu}(p)$ of Eq.~\eqref{eomofshell} by expanding in the four basis vectors
\begin{eqnarray}
u_0^\mu &=& \frac{p^\mu}{N_0}\ ,\ u_1^\mu =\frac{\epsilon^{\mu\nu\rho\sigma}p_\nu n_\rho(k_{AF})_\sigma}{N_1}\ ,\  u_2^\mu = \frac{\epsilon^{\mu\nu\rho\sigma}p_\nu (u_1)_\rho (k_{AF})_\sigma}{N_2}\ ,\ u_3^\mu = \frac{p^2 k_{AF}^\mu - (p\cdot k_{AF})p^\mu}{N_3}\ . \notag \\
\label{basisvectors}%
\end{eqnarray}
Here, the four-vector $n^\mu$ is an arbitrary, observer-covariant, four-vector with at least one
component perpendicular to the subspace formed by $p^\mu$ and $k_{AF}^\mu$. Note that it is generally not possible to use only a single $n^\mu$ vector to cover all of momentum space due ultimately to the theorem that ``the hair on a sphere cannot be combed'', i.e. it is not possible to find a single, smooth, non-vanishing vector field on a sphere.  This problem exists even in the conventional case in which one tries to construct a set of polarization vectors for the transverse, massless photons. There is always at least one direction in momentum space for which the polarization vectors must be non-smooth. This geometrical impediment to constructing a single, global frame field forces one to choose another external vector $m^\mu$ in a different direction than $n^\mu$ to define the polarization vectors in a small cone. 

The $N_i$ in Eq.~\eqref{basisvectors} are normalization factors, that we choose to be real.
The basis vectors are orthogonal in the sense that $u_i\cdot u_j = 0$ if $i\neq j$. However, $u_0$ and $u_3$ become lightlike if $p^2 =0$ while $u_2$, and $u_3$ become lightlike if $(p\cdot k_{AF})^2 = p^2 k_{AF}^2$. 
Note that this construction completely fails when $p^\mu \propto k_{AF}^\mu$. This is related to the 
existence of singular points on the on-shell energy surfaces there.  We will discuss these singular points in more detail below Eq.~\eqref{normalizations}.

Using the basis in Eq.~\eqref{basisvectors} and noticing that
\begin{subequations}
\begin{eqnarray}
\epsilon^{\alpha\beta\mu\nu}(k_{AF})_{\alpha}p_\beta (u_1)_\nu &=& N_2 u_2^\mu\ , \\
\epsilon^{\alpha\beta\mu\nu}(k_{AF})_{\alpha}p_\beta (u_2)_\nu &=& N_2^{-1}(p^2 k_{AF}^2 - (p\cdot k_{AF})^2) u_1^\mu\ ,
\end{eqnarray}
\end{subequations}
it becomes straightforward to determine the polarization vectors. They are given by
\begin{subequations}
\begin{eqnarray}
e^{(0)\mu}(p) &=& u_0^\mu\ , \\
e^{(3)\mu}(p) &=& u_3^\mu\ , \\
e^{(\pm)\mu}(p) &=& \frac{1}{\sqrt{2}}\left(u_2^\mu \pm iN_2^{-1}\sqrt{(p\cdot k_{AF})^2 - p^2 k_{AF}^2}\;u_1^\mu\right)\ ,
\end{eqnarray}
\label{polarvect}%
\end{subequations}
where the square roots in the expressions for $e^{(\pm)\mu}(p)$ are defined by the conventional principal value. The eigenvalues corresponding to the eigenvectors, as defined in Eq.~\eqref{eomofshell}, are
\begin{subequations}
\begin{eqnarray}
\Lambda_0(p) &=& \frac{1}{\xi}(p^2-\xi m_\gamma^2)\ , \\
\Lambda_3(p) &=& p^2 - m_\gamma^2\ , \\
\Lambda_\pm(p) &=& p^2 - m_\gamma^2 \pm 2\sqrt{(p\cdot k_{AF})^2 - p^2 k_{AF}^2}\ .
\end{eqnarray}
\label{eigenvalues}%
\end{subequations}
These observer-scalar functions of $p^\mu$ and $k_{AF}^\mu$ define the dispersion relations for each of the polarization modes by fixing $\Lambda_\lambda(p) = 0$.  Substituting the resulting solutions $p^0 = E_\lambda(\vec{p})$ into the expressions in Eq.~\eqref{polarvect} thus gives the polarization vectors that solve Eq.~\eqref{eom}. Note that the basis set in Eq.~\eqref{polarvect} is valid in a larger part of momentum space, in the sense that the four-vectors solve the off-shell condition in Eq.~\eqref{eomofshell}. This fact will be convenient for the analyses of microcausality and the Feynman propagator. 

Excluding the hypersurfaces in momentum space where $p^2 = 0$ or $(p\cdot k_{AF})^2 - p^2 k_{AF}^2 = 0$, it is always possible to choose the normalization factors in Eq.~\eqref{basisvectors} such that the polarization vectors in Eq.~\eqref{polarvect} are normalized to $+1$ or $-1$. The resulting values for $N_i$, which we choose to be real, are determined by
\begin{eqnarray}
|N_0|^2 &=& |p^2|\ , \notag \\
|N_1|^2 &=& |p^2((n\cdot k_{AF})^2 - n^2 k_{AF}^2)+n^2(p\cdot k_{AF})^2 + k_{AF}^2 (n\cdot p)^2 - 2(p\cdot k_{AF})(n\cdot p)(n\cdot k_{AF})|\ , \notag \\
|N_2|^2 &=& |(p\cdot k_{AF})^2 - p^2 k_{AF}^2|\ ,\notag \\
|N_3|^2 &=& |p^2 (p^2k_{AF}^2-(p\cdot k_{AF})^2)|\ .
\label{normalizations}%
\end{eqnarray} 

As mentioned previously, there are hypersurfaces in momentum space where the definitions in Eqs.~\eqref{basisvectors}, \eqref{polarvect}, and~\eqref{normalizations} are invalid. First, momenta that satisfy $p^2 = 0$ yield polarization vectors $e^{(0)\mu}(p) \propto e^{(3)\mu}(p)$ that are lightlike. This is not a serious problem since $p^2 = 0$ does not intersect the mass-shell of modes $\lambda=0,3$, because of the nonzero photon mass. Ultimately we only need to be able to define the on-shell polarization vectors, while the off-shell polarization vectors are very convenient, but not necessary. The more interesting, perturbed physical $\lambda = \pm$ states are not problematic at $p^2 = 0$.

More serious singular points occur when $(p\cdot k_{AF})^2 = p^2 k_{AF}^2$, which 
 happens, for example, when $p^\mu \propto k_{AF}^\mu$.  This only becomes an issue if the mass shell of a physical polarization mode intersects the momentum-space hypersurface on which the singular points
 lie. 
 When $k_{AF}^2\leq 0$, the singular hypersurface never intersects the mass shell.
  In the case $k_{AF}^2 > 0$, an intersection occurs for all transverse polarization modes if
\begin{equation}
p^\mu = \varsigma\mathcal{K}^\mu \equiv \varsigma\frac{m_\gamma k_{AF}^\mu}{\sqrt{k_{AF}^2}}
\label{specialmomentum}
\end{equation}
with $\varsigma$ either $1$ or $-1$. At these two momenta, the dispersion relations of the modes $\lambda = 3,+,-$ are solved simultaneously, while $p^\mu = \varsigma \mathcal{K}^\mu$ also satisfies $(p\cdot k_{AF})^2 = p^2 k_{AF}^2$. Furthermore, at $p^\mu = \varsigma \mathcal{K}^\mu$, the physical polarization vectors in Eq.~\eqref{polarvect} all vanish. In fact, the LV term in Eq.~\eqref{eom} also vanishes in these cases, so any polarization vector orthogonal to $p^\mu$ will satisfy the equation of motion there. We will choose to define the polarization vectors at $p^\mu = \varsigma\mathcal{K}^\mu$ by taking some limit $p^0 \rightarrow \varsigma\mathcal{K}^0$ of the off-shell polarization vectors evaluated at $\vec{p} = \varsigma\vec{\mathcal{K}}$. At this value of the spatial momentum, the polarization vectors are given by
\begin{subequations}
\begin{eqnarray}
e^{(3)\mu}(p^0,\varsigma\vec{\mathcal{K}}) &=& \frac{\varepsilon(p^0 - \varsigma\mathcal{K}^0)}{\tilde{N}_3}\left(\varsigma\eta^{\mu 0}|\vec{k}_{AF}| + \eta^{\mu}_{\;\;i}\hat{k}_{AF}^i \frac{\sqrt{k_{AF}^2}p^0}{m_\gamma}\right)\ , \\
e^{(\pm)\mu}(p^0,\varsigma\vec{\mathcal{K}}) &=& \frac{1}{\sqrt{2}|\vec{k}_{AF}|\tilde{N}_1}\left(\epsilon^{\mu 0 \rho \sigma}(\tilde{u}_1)_\rho (k_{AF})_\sigma \pm i\varepsilon(p^0 - \varsigma\mathcal{K}^0)|\vec{k}_{AF}|\tilde{u}_1^\mu\right)\ ,
\end{eqnarray}
\label{singvectors}%
\end{subequations}
where $\varepsilon(x) = x/\sqrt{x^2}$, $\tilde{u}_1^\mu = \epsilon^{\mu 0 \rho \sigma}n_\rho (k_{AF})_\sigma$, and $\tilde{N}_3$ and $\tilde{N}_1$ are normalization factors that satisfy
\begin{subequations}
\begin{eqnarray}
|\tilde{N}_1|^2 &=&(\vec k_{AF} \times \vec n)^2 , \\
|\tilde{N}_3|^2 &=& \left|\vec{k}_{AF}^2 - \frac{(p^0)^2k_{AF}^2}{m_\gamma^2}\right| \ .
\end{eqnarray}
\label{normalizationsspecial}%
\end{subequations} 
Note that these definitions fail for $\vec k_{AF} = \vec 0$.
If this is the case, it is always possible to make a small observer Lorentz transformation to a frame in which 
$\vec k_{AF} \ne \vec 0$ and define the theory there.  This choice breaks manifest observer covariance at the singular point, but this seems unavoidable.

We cannot just put the polarization vectors in Eqs.~\eqref{singvectors} on the mass shell defined by $p^0 = \varsigma\mathcal{K}^0$, since the factor $\varepsilon(p^0 - \varsigma\mathcal{K}^0)$ becomes undefined, however, we are free to pick the positive sign that results from approaching the singular point from the direction $p^0 > \varsigma\mathcal{K}^0$ and use it to make a choice at the singular point.  The price we pay for doing this, is that we lose manifest observer Lorentz covariance along a singular line (on shell this corresponds to the two singular points). This does not cause any issues in the current paper as a complete
basis of polarization vectors at each momentum is all that is required for a covariant field expansion, they need not be continuous through the singular point.
In fact, it is not possible even in the conventional massless photon case to find a smooth set of transverse 
polarization vectors that globally covers momentum space due to the topological obstruction involved in 
``combing the hair on a sphere''.  This means that
manifest observer invariance is never possible as certain choices
have to be made as to how the necessary discontinuities are placed in momentum space.
An example of the physical effect of this obstruction can be observed in Berry's Phase \cite{berry}
in which a helicity state, adiabatically transported through a closed loop in momentum space, picks up a non-trivial phase proportional to the solid angle of the loop.
This would not happen if a globally defined frame field of helicity states was possible.

We can now summarize the orthogonality of the polarization vectors (evaluated at the same four-momentum) by
\begin{equation}
e^{(\lambda)*}(p)\cdot e^{(\lambda')}(p) = g^{\lambda \lambda'}\ ,
\label{fourmomentumorthogonality}
\end{equation}
with
\begin{equation}
g = \left\{\begin{array}{cl}
{\rm diag}({1,-1,-1,-1}) & \quad {\rm for}\ \ p^2 > 0\ \\
{\rm diag}({-1,1,-1,-1}) & \quad {\rm for}\ \ p^2 < 0\ \ {\rm and}\ \ (p\cdot k_{AF})^2 - p^2 k_{AF}^2 > 0 \\
\left(\begin{array}{cc}
\mathds{1}_2&0 \\
0& -{\rm sgn}(u_1^2) \sigma_1 \end{array}\right) & \quad {\rm for}\ \ p^2 < 0\ \ {\rm and}\ \ (p\cdot k_{AF})^2 - p^2 k_{AF}^2 < 0 \\
\end{array}\right.\ .
\label{polarmetric}
\end{equation}
In Eq.~\eqref{polarmetric}, $\mathds{1}_2$ is the $2 \times 2$ unit matrix and $\sigma_1$ is the
usual Pauli matrix
\begin{equation}
\sigma_1 = 
\begin{pmatrix}
\;0\ &\ 1\ \ \\
\;1\ &\ 0\ \
\end{pmatrix}
 .
\end{equation}
The indices $\lambda$ and $\lambda'$ in Eq.~\eqref{fourmomentumorthogonality} label the rows and columns of $g$ and run over $0,3,+,-$ in that order.
At the on-shell singular points in Eq.~\eqref{specialmomentum}, where $(p\cdot k_{AF})^2 - p^2 k_{AF}^2 =0$, the lower-right $2 \times 2$ matrix becomes the negative unit matrix, if we use the definitions in Eqs.~\eqref{singvectors}.

Eq.~\eqref{fourmomentumorthogonality} establishes the orthogonality of polarization vectors that are evaluated at the same four-momentum. Since on shell the polarization vectors of different modes are evaluated at different values of $p^0 = E_\lambda(\vec{p})$, Eq.~\eqref{fourmomentumorthogonality} does not represent an orthogonality relation for on-shell eigenvectors. In Ref.~\cite{don-pat-potting2}, such a relation was derived. With a slight change in normalization relative to this reference, it is given by
\begin{align}
&e^{*(\lambda^\prime)}_\mu(\vec p)\Bigl[
\left( E_\lambda(\vec p) + E_{\lambda'}(\vec p)\right)
\left(\eta^{\mu\nu}-(1-\xi^{-1})\delta_0^\mu \delta_0^\nu\right)\nonumber\\
&\qquad\qquad{}-(1-\xi^{-1})p^i(\delta_i^\mu\delta_0^\nu + \delta_0^\mu\delta_i^\nu)
-2 i k_{AF}^\kappa \epsilon_{\kappa 0}{}^{\mu\nu} \Bigr] e_\nu^{(\lambda)}(\vec p)
=  g^{\lambda \lambda^\prime}\left.\Lambda_\lambda'(p)\right|_{p^0 = E_\lambda} \,.
\label{orthog1}
\end{align}
where $\Lambda_\lambda'(p)$ is the derivative of $\Lambda_\lambda(p)$ with respect to $p^0$. The reason for choosing an alternate normalization will become clear when we perform the quantization. Note that the only relevant $g^{\lambda \lambda^\prime}$ for the on-shell states
is the diagonal one, provided $m_\gamma \ne 0$.

The fact that $\left.\Lambda_\lambda'(p)\right|_{p^0 = E_\lambda}$ indeed corresponds to the normalization in Eq.~\eqref{fourmomentumorthogonality} can easily be seen by considering the $p^0$-derivative of Eq.~\eqref{eomofshell}, which reads
\begin{equation}
S'^\mu_{\;\;\nu}e^{(\lambda)\nu}(p) + S^\mu_{\;\;\nu}e^{'(\lambda)\nu}(p) = \Lambda'_\lambda(p) e^{(\lambda)\mu}(p) + \Lambda_\lambda(p) e^{'(\lambda)\mu}(p)\ ,
\label{eomderivative}
\end{equation}
where the primes denote derivatives with respect to $p^0$. After contracting this equation with $\epsilon^{*(\lambda)}_{\mu}(p)$ and substituting $p^0 = E_\lambda(\vec{p})$ everywhere, the second term on both the left-hand side and the right-hand side vanishes. Inspection of the explicit expression for $S'^\mu_{\;\;\nu}$ then reveals that we have obtained Eq.~\eqref{orthog1} for the case $\lambda = \lambda'$, confirming the factor $\left.\Lambda_\lambda'(p)\right|_{p^0 = E_\lambda}$ in that equation. The derivation of Eq.~\eqref{orthog1} in Ref.~\cite{don-pat-potting2} was done for $E_\lambda(\vec{p}) \neq E_{\lambda'}(\vec{p})$. However, using the fact that Eq.~\eqref{fourmomentumorthogonality} holds for on-shell eigenvectors with degenerate energies, we can use Eq.~\eqref{eomderivative} to show that Eq.~\eqref{orthog1} also holds if $E_\lambda(\vec{p}) = E_{\lambda'}(\vec{p})$.

In a similar way an orthogonality relation for polarization vectors with opposite three-momenta can derived~\cite{don-pat-potting2}. As long as $E_\lambda(\vec{p}) \neq - E_{\lambda'}(-\vec{p})$ it holds that
\begin{align}
&e^{(\lambda^\prime)}_\mu(-\vec p)\Bigl[
\left( E_\lambda(\vec p) - E_{\lambda'}(-\vec p)\right)
\left(\eta^{\mu\nu}-(1-\xi^{-1})\delta_0^\mu \delta_0^\nu\right)\nonumber\\
&\qquad\qquad{}-(1-\xi^{-1})p^i(\delta_i^\mu\delta_0^\nu + \delta_0^\mu\delta_i^\nu)
-2 i k_{AF}^\kappa \epsilon_{\kappa 0}{}^{\mu\nu} \Bigr] e_\nu^{(\lambda)}(\vec p)
=  0\,.
\label{orthog2}
\end{align}
Note that there is no complex conjugate on the
left-side polarization vector in this relation.

\section{Analysis of the dispersion relation}\label{sec:disprel}%

In the previous section we found explicit expressions for the eigenvectors of $S^\mu_{\;\;\nu}$ in Eq.~\eqref{polarvect}. These become the on-shell photon polarization vectors if we substitute for $p^0$ the concordant-frame positive root $E_\lambda(\vec{p})$ of $\Lambda_\lambda(p)$, with $\Lambda_\lambda(p)$ defined in Eq.~\eqref{eigenvalues}. In this section we investigate the dispersion relations, given by $\Lambda_\lambda(p) =0$, and in particular the reality, degeneracy, and positivity of their roots \cite{relationtob}.

The full dispersion relation of the CPT-odd photons is given by $\det(S) = \prod_{\lambda}\Lambda_\lambda(p) = 0$, and thus by
\begin{equation}
\frac{1}{\xi}(p^2-\xi m_\gamma^2)(p^2-m_\gamma^2)((p^2-m_\gamma^2)^2 - 4(p\cdot k_{AF})^2 + 4p^2 k_{AF}^2) = 0\ .
\label{disprel}
\end{equation}
The left-hand side is an eighth order polynomial in $p^0$ and as such has eight (possibly complex and/or degenerate) roots, which we label by $\omega_1,\ldots,\omega_8$. Because there is no term proportional to the seventh power of $p^0$ in Eq.~\eqref{disprel}, Vieta's formulas tell us that the sum of all roots vanishes, i.e.
\begin{equation}
\sum_{i=1}^8\omega_i = 0\ .
\end{equation}
The polynomial in Eq.~\eqref{disprel} can be factorized in three separate polynomials, two of which are $\Lambda_0(p)$ and $\Lambda_3(p)$, while the third one is given by
\begin{eqnarray}
\Lambda_T(p) & = & \Lambda_+(p)\Lambda_-(p) = (p^2-m_\gamma^2)^2 - 4(p\cdot k_{AF})^2 + 4p^2 k_{AF}^2\ .
\label{transversedisprel}
\end{eqnarray}

Since Eq.~\eqref{disprel} is invariant under $p \rightarrow -p$, all roots come in pairs such as $\omega_1(\vec{p}) = -\omega_2(-\vec{p})$. In concordant frames, one root of each pair is positive, while the other is negative, e.g. if $\omega_1(\vec{p}) > 0$, then $\omega_2(\vec{p}) < 0$. We apply the usual redefinition to the concordant-frame negative-energy solutions, i.e.
\begin{subequations}
\begin{eqnarray}
E_0(\vec{p}) &=& \omega_1(\vec{p}) =  -\omega_2(-\vec{p}) = \sqrt{\vec{p}^2 + \xi m_\gamma^2}\ , \\
E_3(\vec{p}) &=& \omega_3(\vec{p}) =  -\omega_4(-\vec{p}) = \sqrt{\vec{p}^2 + m_\gamma^2}\ , \\
E_+(\vec{p}) &=& \omega_5(\vec{p}) =  -\omega_6(-\vec{p})\ , \\
E_-(\vec{p}) &=& \omega_7(\vec{p}) =  -\omega_8(-\vec{p})\ .
\end{eqnarray}
\label{energies}%
\end{subequations}
This also defines our labeling of the roots of the different $\Lambda_\lambda(p)$ functions. Although it is not evident at this point, two of the roots ($\omega_5$ and $\omega_6$) correspond to the polarization mode $\lambda = +$, while the other two ($\omega_7$ and $\omega_8$) belong to $\lambda = -$. Together they are the roots of the fourth-order polynomial $\Lambda_T(p)$ in Eq.~\eqref{transversedisprel} and obey
\begin{equation}
\sum_{i=5}^8 \omega_i = 0\ .
\end{equation} 
Since the roots of $\Lambda_0(p)$ and $\Lambda_3(p)$ are trivial and need no further discussion, we will commit the rest of this section to $\Lambda_+(p)$ and $\Lambda_-(p)$.

When not restricting to an observer frame where $k^\mu_{AF}$ has a convenient form, the explicit expressions for the roots of $\Lambda_\pm(p)$ are unwieldy and provide little insight about the issues we want to discuss (except for lightlike $k_{AF}^\mu$). However, even without such explicit expressions, it is possible to show that if $k_{AF}^2 <  m_\gamma^2$, then $\Lambda_+(p)$ and $\Lambda_-(p)$ have two real and non-degenerate roots each. Moreover, these four roots are all different, except at two points in momentum space for timelike $k_{AF}^\mu$. Similarly, we can show that the energy is bounded from below and that no negative energies occur if $(k^0_{AF})^2 < m_\gamma^2$.

\begin{figure}[p]
	\centering
	\includegraphics[width=\textwidth]{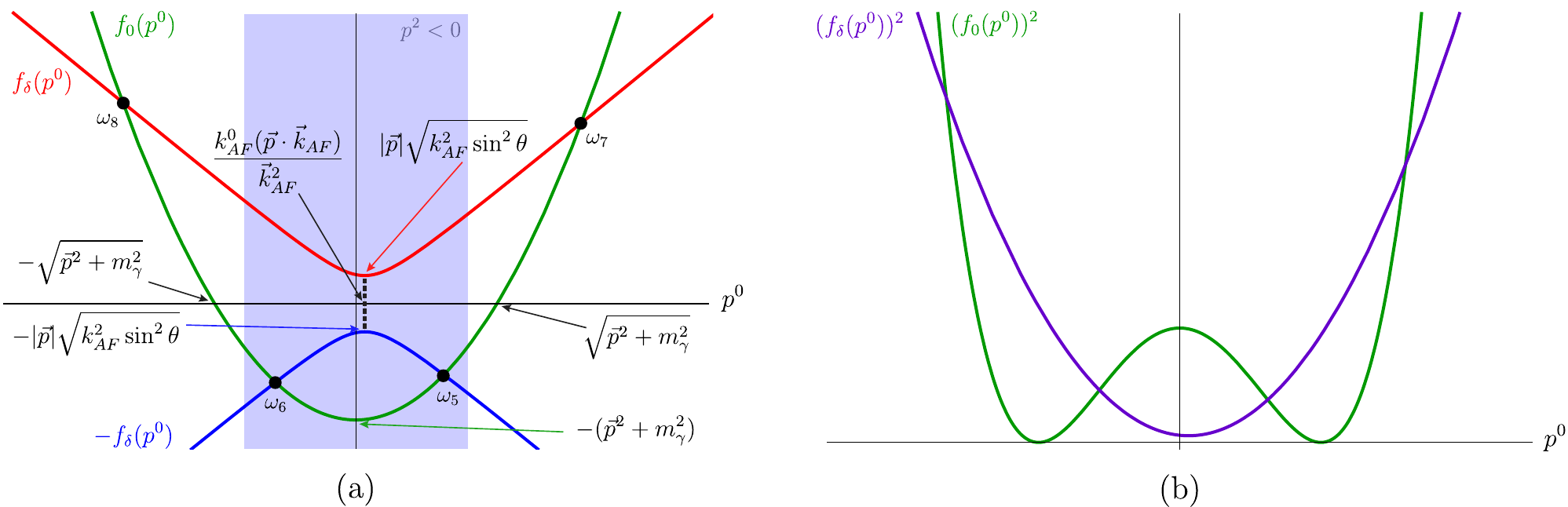}
	\caption{(a) Typical plot of $f_0(p^0)$ and $\pm f_\delta(p^0)$ for timelike $k_{AF}$. The plots are exaggerated in the sense that for physically viable values of $k^\mu_{AF}$ in concordant frames and for experimentally attainable values of $\vec{p}$, both $f_\delta(p^0)$ and $-f_\delta(p^0)$ are nearly horizontal and very close to the $p^0$-axis. Black arrows indicate $p^0$ values, colored arrows indicate values of the corresponding function. (b) Corresponding plot for $(f_0(p^0))^2$ and $(f_\delta(p^0))^2$. The latter stays above the $p^0$ axis, which corresponds to the square root being always real.}
	\label{fig:timelikek}
\end{figure}

To prove the statements in the previous paragraph, we define the following functions of $p^0$:
\begin{subequations}
\begin{eqnarray}
f_0(p^0) &=& \frac{1}{2}\left(\Lambda_+(p^0) + \Lambda_-(p^0)\right)\ , \\
f_\delta (p^0) &=& \frac{1}{2}\left(\Lambda_+(p^0) - \Lambda_-(p^0)\right)\ ,
\end{eqnarray}
\label{partfunctions}%
\end{subequations}
where we now view $\Lambda_\pm(p)$ as functions of $p^0$ by considering them at fixed $\vec{p}$. It follows that $\Lambda_\pm(p^0) = f_0(p^0) \pm f_\delta(p^0)$ and that for a root, $\omega$, of $\Lambda_\pm(p^0)$, $f_0(\omega) = \mp f_\delta(\omega)$, i.e. in a plot the intersections of $f_0(p^0)$ with $\mp f_\delta(p^0)$ correspond to the roots of $\Lambda_\pm(p)$, while intersections of $(f_0(p^0))^2$ and $(f_\delta(p^0))^2$ correspond to the roots of the polynomial $\Lambda_T(p)$. Examples of such plots are given in Figs.~\ref{fig:timelikek} and \ref{fig:spacelikek}.

The derivatives with respect to $p^0$ of the functions in Eq.~\eqref{partfunctions} are given by
\begin{subequations}
\begin{eqnarray}
f'_0(p^0) &=& 2p^0 \stackrel{p^0 \rightarrow \infty}{\longrightarrow} \infty\ , \\
f'_\delta (p^0) &=& \frac{4\vec{k}_{AF}^2 p^0 - 4k_{AF}^0 (\vec{p}\cdot \vec{k}_{AF})}{f_\delta(p^0)} \stackrel{p^0 \rightarrow \infty}{\longrightarrow} 2|\vec{k}_{AF}|\ , 
\end{eqnarray}
\label{derivatives}%
\end{subequations}
where the limiting values are for $p^0$ to positive infinity. Taking $p^0$ to negative infinity will give the same result with opposite sign. The derivative $f'_\delta(p^0)$ shows that if $f_\delta(p^0) \in \mathds{R}$, then $f_\delta(p^0)$ is an increasing (decreasing) function for $p^0$ larger (smaller) than $k^0_{AF}(\vec{p}\cdot\vec{k}_{AF})/\vec{k}_{AF}^2$. To analyze the functions further, we will make a distinction between $k_{AF}^2 \leq 0$ and $k_{AF}^2 > 0$. In the following, we will discuss these timelike and spacelike/lightlike cases separately.

{\it Timelike case --} If $k_{AF}^2 > 0$, a typical plot of the functions $f_0(p^0)$ and $\pm f_\delta(p^0)$ looks like the plot in Fig~\ref{fig:timelikek}(a). The corresponding plots of $(f_0(p^0))^2$ and $(f_\delta (p^0))^2$ are shown in figure Fig~\ref{fig:timelikek}(b). From the limiting values of the derivatives in Eqs.~\eqref{derivatives}, together with the fact that $f_\delta(p^0)$ is real and non-negative if $k_{AF}^2 > 0$, it is easily seen that $f_\delta(p^0)$ always intersects $f_0(p^0)$ at two different points. These points correspond to the two roots of $\Lambda_-(p)$: $\omega_7$ and $\omega_8$. This thus establishes that $\Lambda_-(p)$ always has two non-degenerate roots if $k_{AF}^2 > 0$. One of these roots is positive, while the other one is negative and these signs are the same in any observer frame. Moreover, $E_-(\vec{p})$ is bounded from below, as shown in Eq.~\eqref{lowerboundEmin}.

In Fig~\ref{fig:timelikek}(a), $-f_\delta(p^0)$ also intersects $f_0(p^0)$ twice, once for positive $p^0$ and once for negative $p^0$. However, there are two other possible scenarios. The intersections can be on the same side of the vertical $p^0 = 0$ axis, or the curve of $-f_\delta(p^0)$ might lie entirely below the one of $f_0(p^0)$. In the former case, the roots of $\Lambda_+(p)$ have the same sign, while in the latter case they both have a non-vanishing imaginary part. These three scenarios are summarized by
\begin{equation}
\begin{array}{ccccccc}
(i) & -f_\delta(0) > f_0(0) & & & \rightarrow &\omega_5,\omega_6 \in \mathds{R}\ , & {\rm sgn}(\omega_5) = - {\rm sgn}(\omega_6)\ , \\
(ii) & -f_\delta(0) < f_0(0) &{\rm and} & \exists p^0: -f_\delta(p^0) > f_0(p^0) & \rightarrow & \omega_5,\omega_6 \in \mathds{R}\ , & {\rm sgn}(\omega_5) = {\rm sgn}(\omega_6)\ , \\
(iii) & & & \forall p^0: -f_\delta(p^0) < f_0(p^0) & \rightarrow & \omega_5,\omega_6 \in \mathds{C}\ . &
\end{array} \notag \\
\end{equation}
It turns out that a sufficient observer non-invariant condition for scenario $(i)$ is 
\begin{equation}
(k^0_{AF})^2 < m_\gamma^2\ .
\label{normalsol}%
\end{equation}
The fact that $-f_\delta(0) > f_0(0)$ if this condition holds is shown in Appendix~\ref{appA}. We also find there that, if $(k^0_{AF})^2 > m_\gamma^2$, then there exist a range of (generally small) three-momenta for which $-f_\delta(0) < f_0(0)$, so either scenario $(ii)$ or $(iii)$ applies.

In Appendix~\ref{appA} we show that if the observer Lorentz invariant condition
\begin{equation}
k_{AF}^2 < m_\gamma^2
\label{normalsol2}
\end{equation}
holds, we can always find a $p^0$ for which $-f_\delta(p^0) > f_0(p^0)$. This shows that both roots of $\Lambda_+(p)$ are real and non degenerate if Eq.~\eqref{normalsol2} is satisfied, while their sign is guaranteed to differ if Eq.~\eqref{normalsol} applies. Eq.~\eqref{normalsol2} can be enforced on the theory in any observer frame. However, the size of $k^0_{AF}$ changes when performing an observer Lorentz boost. Therefore it can only be satisfied in a subset of frames, which we can call concordant frames. In other words, Eq.~\eqref{normalsol} provides a quantitative definition of a concordant frame in the case where $k^\mu_{AF}$ is the only Lorentz-violating coefficient. The fact that such a definition is possible hinges on the introduction of a nonzero photon mass. 

In non-concordant frames, the signs of the two roots of $\Lambda_+(p)$ can thus be equal. If they are both negative, the energy (given in Eq.~\eqref{energies}) is also negative. However, since this only happens for a limited range of $|\vec{p}|$ values (see Eq.~\eqref{appcond3}), $E_+(\vec{p})$ must be bounded from below. In fact, in Eq.~\eqref{lowerboundEplus}, we determine that $E_+(\vec{p}) \geq \sqrt{m_\gamma^2 - k_{AF}^2} -|\vec{k}_{AF}|$.

If Eq.~\eqref{normalsol2} is satisfied, the only degeneracy in the dispersion relation for the $\lambda = \pm$ modes can come from a root of $\Lambda_-(p)$ being equal to a root of $\Lambda_+(p)$. This requires $f_\delta(p^0) = 0$ while $p^0$ simultaneously has to solve $p^2 = m_\gamma^2$. It follows that the roots of $\Lambda_-(p)$ and $\Lambda_+(p)$ become equal if 
\begin{equation}
p^\mu = \varsigma \frac{m_\gamma k_{AF}^\mu}{\sqrt{k^2}} \equiv \varsigma \mathcal{K}^\mu\ ,
\end{equation}
which are points in momentum space where the LV term disappears from the equation of motion, as already discussed in Section~\ref{sec:polvec}.

{\it Spacelike/lightlike case --} If $k_{AF}^2 \leq 0$, a typical plot of the functions in Eqs.~\eqref{partfunctions} looks like the one in Fig.~\ref{fig:spacelikek}(a). One clearly sees that the square root in $f_\delta(p^0)$ becomes imaginary for values of $p^0$ between $x_{-1}$ and $x_{+1}$, with
\begin{equation}
x_\alpha = \frac{k_{AF}^0(\vec{p}\cdot\vec{k}_{AF})}{\vec{k}_{AF}^2} +  \frac{\alpha\sqrt{k_{AF}^2\left((\vec{p}\cdot\vec{k}_{AF})^2-\vec{p}^2\vec{k}_{AF}^2\right)}}{\vec{k}_{AF}^2}\ .
\label{xplusmin}%
\end{equation}%
However, we show in Appendix~\ref{appA} that $|x_\alpha| < \sqrt{\vec{p}^2 + m_\gamma^2}$ for all values of $\vec{p}$ and $k^\mu_{AF}$. This means that $\Lambda_+(p)$ and $\Lambda_-(p)$ always have two real roots each, if $k^\mu_{AF}$ is spacelike or lightlike. As in the case of timelike $k^\mu_{AF}$, we find, by investigating when $-f_\delta(0) < f_0(0)$ (see Appendix~\ref{appA}), that the condition in Eq.~\eqref{normalsol} is sufficient to make sure that the roots of $\Lambda_+(p)$ have opposite signs. On the other hand, if $(k^0_{AF})^2 > m_\gamma^2$ then there exist observer frames in which both roots have the same sign. As in the timelike case, $E_+(\vec{p})$ can thus become negative, however in appendix~\ref{appB} we show that $E_\pm(\vec{p}) \geq \sqrt{m_\gamma^2 - k_{AF}^2} \mp |\vec{k}_{AF}|$ for spacelike/lightlike $k_{AF}^\mu$.

We summarize our findings regarding the signs and the domain of the roots in Table~\ref{Tab1}. It is clear that $k_{AF}^2 < m_\gamma^2$ is a necessary condition for a consistent physical theory in all observer frames. This shows that introducing a nonzero photon mass is unavoidable if $k^\mu_{AF}$ is timelike. This was already found in Ref.~\cite{alfaro-cambiaso}. In addition, both for spacelike, lightlike, and timelike $k^\mu_{AF}$, a nonzero photon mass allows for a quantitative definition of concordant frames, in the sense that in frames where Eq.~\eqref{normalsol} is satisfied, energy positivity is guaranteed (some additional issues related to energy positivity are discussed in Section~\ref{sec:energypositivity}).

\begin{table}[t]
\centering
\setlength{\tabcolsep}{10pt}
\begin{tabular}{|ccc|cc|}
\hline
$k_{AF}^2$ & $(k^0_{AF})^2$ & ${\rm sgn}(k^0_{AF}(\vec{p}\cdot\vec{k}_{AF}))$ & {\rm domain\ roots} & ${\rm sgn}(\omega_5,\omega_6,\omega_7,\omega_8)$  \\
\hline
$k_{AF}^2 < 0$ & $< m_\gamma^2$ & $+$ or $-$ & $\mathds{R}$ & $(+,-,+,-)$ \\
$k_{AF}^2 < 0$ & $> m_\gamma^2$ & $+$ & $\mathds{R}$ & $(+,-,+,-)$ or $(+,+,+,-)$ \\
$k_{AF}^2 < 0$ & $> m_\gamma^2$ & $-$ & $\mathds{R}$ & $(+,-,+,-)$ or $(-,-,+,-)$ \\
$0 < k_{AF}^2 < m_\gamma^2$ & $< m_\gamma^2$ & $+$ or $-$ & $\mathds{R}$ & $(+,-,+,-)$ \\
$0 < k_{AF}^2 < m_\gamma^2$ & $> m_\gamma^2$ & $+$ & $\mathds{R}$ & $(+,-,+,-)$ or $(+,+,+,-)$ \\
$0 < k_{AF}^2 < m_\gamma^2$ & $> m_\gamma^2$ & $-$ & $\mathds{R}$ & $(+,-,+,-)$ or $(-,-,+,-)$ \\
$k_{AF}^2 > m_\gamma^2$ & $> m_\gamma^2$ & $+$ or $-$ & $\mathds{C}$ & n.a. \\
\hline\hline
\end{tabular}
\caption{The different conditions on $k^\mu_{AF}$ in the three columns on the left give different possibilities for the sign and domain of the roots of the $\lambda = \pm$ dispersion relations. The latter are summarized in the two right-most columns. Which of the options in the right-most column is realized, is determined by $|\vec{p}|$ and the angle between $\vec{p}$ and $\vec{k}_{AF}$. If $|\vec{p}|$ is in the interval in Eq.~\eqref{appcond3} and the angle satisfies Eq.~\eqref{appcond3b}, then three of the four roots will have the same sign (provided $(k^0_{AF})^2 > m_\gamma^2$).}
\label{Tab1}
\end{table}

It is interesting, therefore, to compare the current experimental bounds on $k^\mu_{AF}$ and $m_\gamma$.
For the photon mass, the particle data group (PDG) quotes as the best possible bound \cite{pdg}
\begin{equation}
m_\gamma < 1 \times 10^{-27}\,\mbox{GeV}.
\label{photonmassbound}
\end{equation}
This limit is inferred from the absence of a perturbed structure of large-scale
magnetic fields that would result from a significant nonzero photon mass (see Refs.~\cite{ryutov,goldhaber}).
The verification of certain properties of galactic magnetic fields might allow for an improvement
of Eq.~\eqref{photonmassbound} by nine orders of magnitude \cite{chibisov,goldhaber}.
Nevertheless, this result is still many orders of magnitude from the best bounds on $k^\mu_{AF}$, which follow from cosmological searches for birefringence \cite{datatables}:
\begin{equation}
k_{AF}^0 < 10^{-43}\,\mbox{GeV},
\end{equation}
where $k_{AF}^0$ is defined in Sun-centered inertial reference frame \cite{datatables}.
It follows that the assumption of a nonzero photon mass, that permits the construction of a phenomenologically viable model
for photons with CPT-violation (with $k_{AF}^2 < (k_{AF}^0)^2 < m_\gamma^2$), is entirely consistent with experimental observations.  Moreover,
the St\"uckelberg mechanism can be used to introduce the mass in a gauge-invarant manner, 
at least at the level of pure QED.

This does not mean, however, that we can always ignore the negative energies in the theory, even for practical purposes. This is illustrated for example by assuming that the sizes of the photon mass and $k^\mu_{AF}$ are comparable with the best achievable bounds, such that $m_\gamma \sim 10^{-36}$ GeV, as mentioned below Eq.~\eqref{photonmassbound}. Frames moving with respect to Earth with a relativistic $\gamma$-factor up to $\gamma \sim {m_\gamma / k_{AF}^0} \sim 10^7$ can then be considered to be concordant frames (i.e. there are no negative energies in these frames). Inversely, this means that the rest frame of ultra-high-energy cosmic-ray protons could
easily be a non-concordant frame, since these protons have energies up to $10^8$ TeV, corresponding to $\gamma = 10^{11}$. Note that the bound $m_\gamma < 10^{-36}$ GeV is quite speculative \cite{goldhaber},
since it depends on several assumptions about galactic magnetic fields. Taking the PDG value in Eq.~\eqref{photonmassbound} for the photon mass avoids any potential problems with non-concordant frames.
These values are discussed further in the context of Cherenkov radiation in Ref.~\cite{don-pat-potting2}.

\section{Energy positivity and stability}\label{sec:energypositivity}

In the previous section we found that there exist (strongly boosted, but possibly physically relevant) observer frames in which the $\lambda = +$ polarization mode of the LV photon has negative energies for a certain range of three momenta (see Eq.~\eqref{appcond3} in Appendix~\ref{appA}). Nevertheless, the energy remains bounded from below. 

It lies outside the scope of this paper to rigorously address if such a theory can be fully consistent, however, the point of view one often takes in this respect is to regard the theory as an effective theory. The effective theory is only valid up to a certain energy scale, or, equivalently, describes particles restricted to have concordant rest frames. Above this energy scale, unknown higher-dimensional nonrenormalizable operators become relevant. These are conjectured to prevent negative energies in all observer frames. This is discussed in detail in Ref.~\cite{kostelecky-lehnert}.

It was also noticed in Ref.~\cite{kostelecky-lehnert} that negative-energy issues are closely related to the stability of the theory. Photons with the $\lambda = +$ polarization mode can have spacelike momenta and can thus be emitted by an electron or positron traveling fast enough in vacuum. Since an appropriate Lorentz boost of a spacelike momentum can change the sign of its zeroth component, this corresponds to the existence of negative energies in some frame. Although already discussed in Ref.~\cite{kostelecky-lehnert}, we emphasize once again that allowing for vacuum-Cherenkov radiation in an observer Lorentz invariant theory is thus equivalent to accepting that the theory has negative energies in some frame. The alternative of assuming the existence of nonrenormalizable, higher-order operators that prevent negative energies in all observer frames, also prevents vacuum-Cherenkov radiation in nature. 

A common problem in generic theories with negative energies seems to be that there is no simple separation between the positive- and negative-energy branches of the dispersion relation. This impedes the canonical quantization of the theory.
In a Lorentz-symmetric theory, one can easily select one of the branches by using the sign of the roots, which is invariant under Lorentz transformations if all on-shell momenta are timelike. Since the model with $k_{AF}^\mu$ that we consider here, allows for spacelike momenta and negative energies, the sign of the root can no longer be used to select a particular branch of the dispersion relation. It turns out, however, that a function still exists whose sign, when evaluated at a root of $\Lambda_\lambda(p)$, is the same in any observer frame. It is given by
\begin{equation}
\Lambda'_\lambda(p) = \frac{\partial \Lambda_\lambda(p)}{\partial p^0}\ .
\end{equation}
This function can therefore be used to separate the branches in an observer-covariant way even
when a single branch dips into the negative energy region.
The key fact that makes this work is that the ordering of the roots of the dispersion relation is observer Lorentz invariant.

\begin{figure}[p]
	\centering
	\includegraphics[width=\textwidth]{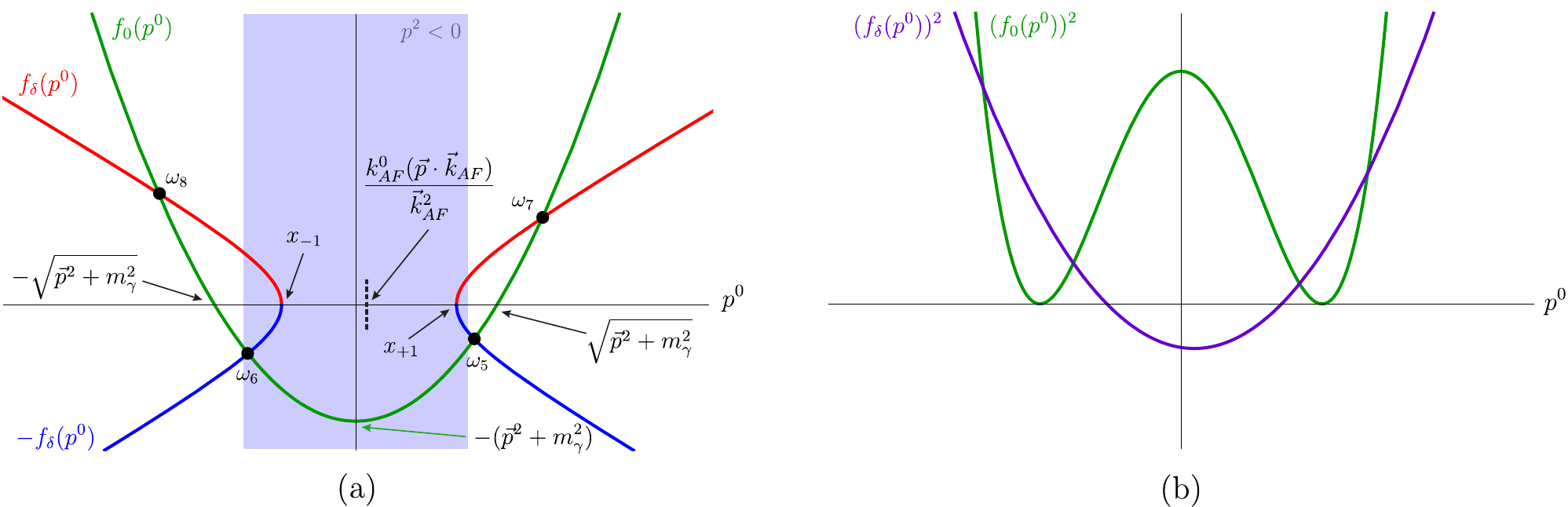}
	\caption{(a) Typical plot of $f_0(p^0)$ and $\pm f_\delta(p^0)$ for spacelike or lightlike $k^\mu_{AF}$. The plots are exaggerated in the sense that for physically viable values of $k^\mu_{AF}$ in concordant frames and for experimentally attainable values of $\vec{p}$, the two branches of both $f_\delta(p^0)$ and $-f_\delta(p^0)$ are nearly horizontal and very close to the $p^0$ axis, while their starting points are also very close together. Black arrows indicate $p^0$ values, colored arrows indicate values of the corresponding function. (b) Corresponding plot for $(f_0(p^0))^2$ and $(f_\delta(p^0))^2$. The latter goes below the $p^0$ axis, which corresponds to the square root becoming imaginary.}
	\label{fig:spacelikek}
\end{figure}

To prove that this function indeed has the correct properties, we first label the roots of $\Lambda_T(p) = \Lambda_+(p)\Lambda_-(p)$ as before, such that $\omega_5$ and $\omega_7$ are the roots on the right-hand side in Figs.~\ref{fig:timelikek} and \ref{fig:spacelikek} (these are positive in concordant frames), while $\omega_6$ and $\omega_8$ are the roots on the left-hand side in the same figures (these are negative in concordant frames). We observe that we can write
\begin{equation}
\Lambda'_T(\omega_j,\vec{p}) = \left[\frac{\partial}{\partial p^0}\prod_{i=5}^8(p^0 - \omega_i)\right]_{p^0 = \omega_j} = \prod_{i\neq j}(\omega_j - \omega_i) \ .
\end{equation}
By inspection of Figs.~\ref{fig:timelikek} and \ref{fig:spacelikek}, together with the considerations in Appendix~\ref{appA}, it is not hard to establish that 
\begin{subequations}
\begin{eqnarray}
\Lambda'_T(\omega_{6},\vec{p}) , \Lambda'_T(\omega_{7},\vec{p}) > 0 \ , \\
\Lambda'_T(\omega_{5},\vec{p}) , \Lambda'_T(\omega_{8},\vec{p}) < 0 \ .
\end{eqnarray}
\label{LambdaTsigns}%
\end{subequations}
For example, 
\begin{equation}
\Lambda'_T(\omega_5,\vec{p}) = (\omega_5 - \omega_6)(\omega_5 - \omega_7)(\omega_5 - \omega_8)\ .
\label{transverseprime}%
\end{equation}
From Figs.~\ref{fig:timelikek} and \ref{fig:spacelikek} it is easy to see that $|\omega_7| > |\omega_5|$, $|\omega_8| > |\omega_5|$, $\omega_7 > 0$, and $\omega_8 < 0$, except at $p^\mu = \varsigma \mathcal{K}^\mu$ for timelike $k^\mu_{AF}$ (see Eq.~\eqref{specialmomentum}), where $\omega_5 = \omega_7 = -\omega_8$. If the roots are not degenerate, the product of the last two factors in Eq.~\eqref{transverseprime} is smaller than zero. Moreover, $\omega_5 - \omega_6$ is also always larger than zero, because the ordering of these roots is the same in any observer frame. This follows directly from considerations in Appendix~\ref{appA}. We conclude that $\Lambda'_T(\omega_5,\vec{p}) < 0$. The sign of $\Lambda'_T(\omega_j,\vec{p})$ evaluated at the other three roots is determined similarly and the result corresponds to Eq.~\eqref{LambdaTsigns}. 

Subsequently, we note that
\begin{subequations}
\begin{eqnarray}
\Lambda'_T(\omega_{5,6},\vec{p}) &=& \Lambda_-(\omega_{5,6},\vec{p})\Lambda'_+(\omega_{5,6},\vec{p})\ , \\
\Lambda'_T(\omega_{7,8},\vec{p}) &=& \Lambda_+(\omega_{7,8},\vec{p})\Lambda'_-(\omega_{7,8},\vec{p})\ .
\end{eqnarray}
\end{subequations}
Examination of Figs.~\ref{fig:timelikek} and \ref{fig:spacelikek} reveals that $\Lambda_-(\omega_{5,6},\vec{p}) < 0$ and $\Lambda_+(\omega_{7,8},\vec{p}) > 0$. Combining this with Eq.~\eqref{LambdaTsigns}, we conclude that
\begin{subequations}
\begin{eqnarray}
\Lambda'_+(\omega_{5},\vec{p}) , \Lambda'_-(\omega_{7},\vec{p}) > 0 \ , \\
\Lambda'_+(\omega_{6},\vec{p}) , \Lambda'_-(\omega_{8},\vec{p}) < 0 \ .
\end{eqnarray}
\label{derivresults}%
\end{subequations}

At $p^\mu = \varsigma \mathcal{K}^\mu$, the expression in Eq.~\eqref{transverseprime} vanishes and the derivation of Eqs.~\eqref{derivresults} fails. As discussed before, the LV term disappears from the equation of motion in that case and at $\vec{p} = \varsigma\vec{\mathcal{K}}$ we have
\begin{equation}
\Lambda_\pm^\prime(p^0, \varsigma\vec{\mathcal{K}}) = 2p^0 \pm 2\varepsilon(p^0 - \varsigma\mathcal{K}^0)|\vec{k}_{AF}|\ ,%
\end{equation}%
with $\varepsilon(x) = x/\sqrt{x^2}$. At $p^0 = \varsigma\mathcal{K}^0$, $\varepsilon(p^0 - \varsigma\mathcal{K}^0)$ is undefined. However, it can be defined using the same limiting procedure that was used for 
the polarization vectors in Eq.~\eqref{singvectors} with $\varepsilon(p^0 - \varsigma\mathcal{K}^0) = 1$ and thus $\left.{\rm sgn}(\Lambda_\pm^\prime(p^0, \varsigma\vec{\mathcal{K}}))\right|_{p^0\rightarrow \varsigma\mathcal{K}^0} = {\rm sgn}\left(\varsigma\mathcal{K}^0\right)$, because $|\mathcal{K}^0| > |\vec{k}_{AF}|$ for timelike $k^\mu_{AF}$. Therefore, also in the degenerate case, the sign of $\Lambda'_\pm(p)$ is an observer Lorentz invariant quantity. In fact, it corresponds to the sign of $p^0$.

From this and from Eqs.~\eqref{derivresults} we thus conclude that the sign of $\Lambda'_\pm(p)$, evaluated at one of its roots, corresponds to the sign of that root in a concordant frame and is an observer Lorentz invariant quantity. This obviously holds for all functions $\Lambda_\lambda(p)$, since for the other polarization modes $\Lambda'_{0,3}(p) \propto p^0$. The fact that it also holds for the polarization modes $\lambda = \pm$ is directly related to the fact that the ordering of the roots stays the same in any observer frame (provided $k_{AF}^2 < m_\gamma^2$), as becomes clear from the considerations below Eq.~\eqref{transverseprime}.

More insight as to why the sign of $\Lambda'_\lambda(p)$ is an observer Lorentz invariant quantity can be gained from considering the group velocity, defined by
\begin{equation}
\vec{v}^{(\lambda)}_g = \frac{\partial E_\lambda(\vec{p})}{\partial \vec{p}}\ .
\label{groupvelocity}
\end{equation}
The size of $\vec{v}^{(\pm)}_g$ is related to the sign of $\Lambda'_\pm(p)$.
To show this, we perform an observer-Lorentz-transformation on $\Lambda'_\lambda(p)$ and obtain
\begin{equation}
\left.\frac{\partial \Lambda_\lambda(p)}{\partial p^0}\right|_{p^0 = E_\lambda(\vec{p})} \longrightarrow \gamma \left[\frac{\partial \Lambda_\lambda(p)}{\partial p^0} - \vec{\beta}\cdot \frac{\partial \Lambda_\lambda(p)}{\partial \vec{p}}\right]_{p^0 = E_\lambda(\vec{p})} = \gamma \left(1 + \vec{\beta} \cdot \vec{v}^{(\lambda)}_g\right) \left[\frac{\partial \Lambda_\lambda(p)}{\partial p^0}\right]_{p^0 = E_\lambda(\vec{p})}\ ,
\label{nosignchange}%
\end{equation}
where $\gamma = 1/\sqrt{1-\vec{\beta}^2}$ is the relativistic boost factor. We used the fact, clarified in Appendix~\ref{appB}, that $\vec{v}^{(\lambda)}_g = -\left[\frac{\partial \Lambda_\lambda(p)}{\partial \vec{p}}\big/ \frac{\partial \Lambda_\lambda(p)}{\partial p^0}\right]_{p^0 = E_\lambda(\vec{p})}$. It is clear that if $|\vec{v}^{(\lambda)}_g| < 1$, then $\Lambda'_\lambda(p)$ has the same sign in any observer frame. In Appendix~\ref{appB} we show explicitly that $|\vec{v}^{(\lambda)}_g| < 1$ for the present LV model.

The considerations above allow us to use the sign of the function $\Lambda'_\pm(p)$ as an observer-Lorentz-invariant way of selecting a particular branch of the dispersion relation. For example:
\begin{equation}
\int d^3 p \int dp^0 h(p^0) {\rm sgn}\left(\Lambda'_+(p)\right) \theta\left(\Lambda'_+(p)\right)\delta(\Lambda_+(p)) = \int \frac{d^3 p}{\Lambda'_+(\omega_5)}h(\omega_5)\ ,
\label{selectionfunction}
\end{equation}
where $h(p^0)$ is an arbitrary function of $p^0$. Incidentally, this shows that $\frac{d^3 p}{\Lambda'_\lambda(p)}$ is an observer Lorentz invariant phase-space factor, that can be used to replace the usual $\frac{d^3 p}{2p^0}$, which is used in the Lorentz-symmetric case.

\section{Propagator and polarization vectors}\label{sec:propandpol}

In this section we derive a relation between a sum over bilinears of polarization vectors and the propagator in momentum space. Since a propagator in coordinate space is a Green's function of the equation-of-motion operator, the momentum-space propagator $P^{\mu\nu}(p)$ satisfies
\begin{equation}
S_{\mu\nu}(p)P^{\nu\rho}(p) = -i\delta_{\mu}^{\;\rho}\ ,
\end{equation}
with $S^\mu_{\;\;\nu}(p)$ defined in Eq.~\eqref{eom}. Therefore,
\begin{equation}
P^{\mu\nu}(p) = -i(S^{-1})^{\mu\nu} = -i\left(\frac{{\rm Adj}(S)}{\det(S)}\right)^{\mu\nu}\ .
\label{deltatoS}
\end{equation}
The determinant of $S$ is given on the left-hand side of Eq.~\eqref{disprel}, while the adjugate matrix can be determined explicitly in terms of traces of powers of $S$ \cite{cambiaso-lehnert-potting}. The result is
\begin{eqnarray}
iP^{\mu\nu}(p) &=& \frac{1}{\Lambda_T(p)}\Big[(p^2-m_\gamma^2)\eta^{\mu\nu} + \frac{4\left(p^\mu p^\nu k_{AF}^2 + k_{AF}^\mu k_{AF}^\nu p^2 - (p^\mu k_{AF}^\nu + p^\nu k_{AF}^\mu)(p\cdot k_{AF}) \right)}{p^2-m_\gamma^2} \notag \\
&& + 2i\epsilon^{\mu\nu\alpha\beta}(k_{AF})_\alpha p_\beta\Big] -(1-\xi)\frac{p^\mu p^\nu}{(p^2-\xi m_\gamma^2)(p^2- m_\gamma^2)}\ ,
\label{propagatormomentumspace}
\end{eqnarray}
with $\Lambda_T(p)$ given in Eq.~\eqref{transversedisprel}.

If there are four orthogonal polarization vectors, we can derive a relation between the expression in Eq.~\eqref{propagatormomentumspace} and the polarization vectors as defined in Eq.~\eqref{polarvect}. To show this, we define a matrix $U$ that has the polarization vectors as its columns, i.e. its entries are defined by
\begin{equation}
U^{ab} = e^{(b)a} \qquad a,b \in {0,1,2,3}\ ,
\end{equation}
where we identify $e^{(b)}$ with $e^{(0)}$, $e^{(3)}$, $e^{(+)}$, $e^{(-)}$ for $b=0,1,2,3$ respectively. It is clear that $(SU)^{ab}  = \Lambda_b(p)e^{(b)a}$ (the matrix $S$ here corresponds to $S^\mu_{\;\;\nu}$, with its first index up and its second index down). Assuming that the polarization vectors are normalized according to Eq.~\eqref{normalizations}, we conclude that
\begin{equation}
(U^\dagger \eta S U)^{ab} = \Lambda_b(p) g^{ab}\ ,
\end{equation}
with $\eta$ the Minkowski metric and $g$ given in Eq.~\eqref{polarmetric}. If all the polarization vectors are orthogonal then $U$ has an inverse:
\begin{equation}
(g^{-1} U^\dagger \eta U)^{ab} = \delta^{ab}\ ,
\label{Uinverse}
\end{equation}
and we can write
\begin{equation}
(S^{-1})^{ab} = (U D^{-1} U^\dagger\eta)^{ab}\ ,
\label{Sinverse}
\end{equation}
where $D^{ab} = \Lambda_b(p) g^{ab}$. Writing with Lorentz indices and once again labeling polarizations by $\lambda = 0,3,+,-$, this becomes
\begin{equation}
iP^{\mu}_{\;\;\nu}(p) = (S^{-1})^\mu_{\;\;\nu} = \sum_{\lambda\lambda'}g^{\lambda\lambda'}\frac{e^{(\lambda)\mu}e^{(\lambda')*}_\nu}{\Lambda_\lambda(p)}\ .
\label{PolarProp}
\end{equation} 
This sum, containing bilinears of the polarization vectors, is thus equal to the expression in Eq.~\eqref{propagatormomentumspace}.

Off shell, the form of two of the four terms in Eq.~\eqref{PolarProp} depends on the sign of $(p\cdot k_{AF})^2 - p^2k_{AF}^2$. This follows from the dependence of $g^{\lambda\lambda'}$ on this same sign (see Eq.~\eqref{polarmetric}). We write the relevant terms as
\begin{eqnarray}
P^{\mu\nu}_T(p) &=& \left\{\begin{array}{cc} g^{++}\frac{e^{(+)\mu} e^{(+)\nu *}}{\Lambda_+(p)} + g^{--}\frac{e^{(-)\mu} e^{(-)\nu *}}{\Lambda_-(p)} &\qquad\quad {\rm for}\quad (p\cdot k_{AF})^2 - p^2 k_{AF}^2 > 0 \\
g^{+-}\frac{e^{(+)\mu} e^{(-)\nu *}}{\Lambda_+(p)} + g^{-+}\frac{e^{(-)\mu} e^{(+)\nu *}}{\Lambda_-(p)} &\qquad\quad {\rm for}\quad (p\cdot k_{AF})^2 - p^2 k_{AF}^2 < 0\end{array}\right.\ .
\label{transversepartpol}
\end{eqnarray}
Notice that each of the terms in Eq.~\eqref{transversepartpol} has a branch cut in the complex $p^0$ plane, due to the square root in the expression for $\Lambda_\pm(p)$. This seems to hamper the definition of an appropriate contour integral to implement the boundary conditions of for example the Feynman propagator. However, the expression in Eq.~\eqref{propagatormomentumspace}, and therefore the entire sum in Eq.~\eqref{PolarProp}, has no such branch cuts. In fact, if we put in the explicit expressions for the polarization vectors, $\Lambda_\pm(p)$, and components of $g$, we find that
\begin{eqnarray}
P^{\mu\nu}_T(p) &=& \frac{1}{\Lambda_T(p)}\left[(p^2-m_\gamma^2)\eta^{\mu\nu} + 2i\epsilon^{\mu\nu\alpha\beta}(k_{AF})_\alpha p_\beta\right] \notag \\
&& + \frac{(p^2-m_\gamma^2)(p^\mu p^\nu k_{AF}^2 + k_{AF}^\mu k_{AF}^\nu p^2 - (p^\mu k_{AF}^\nu + p^\nu k_{AF}^\mu)(p\cdot k_{AF}))}{((p\cdot k_{AF})^2-p^2k_{AF}^2)\Lambda_T(p)}\ ,
\label{Ptransv}
\end{eqnarray}
for both positive and negative $(p\cdot k_{AF})^2 - p^2 k_{AF}^2$. This expression has no branch cuts in the complex $p^0$ plane.
Note that the dependence on $n^\mu$ introduced to define the polarization vectors has dropped
out of the above expression.

\section{Quantization}\label{quantization}

Using the polarization vectors that follow from the equation of motion, given in Eq.~\eqref{polarvect}, we can give the explicit mode expansion of the photon field:
\begin{equation}
A_\mu(x) = \int \frac{d^3 \vec{p}}{(2\pi)^3}\sum_\lambda \frac{1}{\Lambda'_{\lambda}(p)}\left[a_{\vec{p}}^\lambda\; e_\mu^{(\lambda)}(\vec{p})e^{-ip\cdot x} + a_{\vec{p}}^{\lambda\dagger}\; e_\mu^{(\lambda)*}(\vec{p})e^{ip\cdot x}\right]_{p^0=E_\lambda(\vec{p})}\ ,
\label{fouriermodes}
\end{equation}
where $\Lambda'_{\lambda}(p)$ is the derivative with respect to $p^0$ of $\Lambda_\lambda(p)$, defined in Eq.~\eqref{eigenvalues}. This normalization differs from the conventional one and corresponds to the one chosen in Eq.~\eqref{orthog1}. As mentioned below Eq.~\eqref{selectionfunction}, in this way the phase space factor in this expression for $A_\mu(x)$ is observer Lorentz invariant. The complex weights $a_{\vec{p}}^\lambda$ and $a_{\vec{p}}^{\lambda\dagger}$ become annihilation and creation operators on a Fock space, when we quantize the theory.

To perform the quantization, we compute the canonical conjugate of $A_\mu(x)$ in the usual way by taking derivatives of the Lagrangian with respect to the time derivative of the photon field. This results in a canonical momentum, given by
\begin{equation}
\pi^\mu(x) = F^{\mu 0}(x) + \epsilon^{0\mu\alpha\beta}(k_{AF})_\alpha A_\beta(x) - \eta^{\mu 0}\frac{1}{\xi}\partial_\nu A^\nu(x)\ .
\label{canonicalmomentum}
\end{equation}
We then impose the following equal-time commutation relations on the fields:
\begin{subequations}
\begin{eqnarray}
\left[A_\mu(t,\vec{x}),\pi^\nu(t,\vec{y})\right] &=& i\delta^{\nu}_\mu \delta^3(\vec{x}-\vec{y})\ , \\
\left[A_\mu(t,\vec{x}),A_\nu(t,\vec{y})\right] &=& 0 . \label{comrels1}
\end{eqnarray}
\end{subequations}
This implements the standard canonical quantization in a covariant manner, as is done in the conventional Gupta-Bleuler method. From the imposed commutation relations and the expression for the canonical momentum, we find the following commutation relations \cite{don-pat-potting2}:
\begin{subequations}
\begin{align}
[\dot A^\mu(t,\vec x),A^\nu(t,\vec y)]&= -[A^\mu(t,\vec x),\dot A^\nu(t,\vec y)]
=
i\left(\eta^{\mu\nu}-\delta^\mu_0\delta^\nu_0(1-\xi)\right)
\delta^3(\vec x-\vec y)\,,
\label{comrels2}\\
[\dot A_\mu(t,\vec x),\dot A_\nu(t,\vec y)]&=
i\left[2\epsilon^{0\mu\nu\lambda}(k_{AF})_\lambda
+(1-\xi)\left(\delta^\mu_0\delta^\nu_j+\delta^\mu_j\delta^\nu_0\right)
\partial_x^j\right]\delta^3(\vec x-\vec y)\,. 
\label{comrels3}
\end{align}
\label{comrels}%
\end{subequations}

In order to see what the commutation relations in Eqs.~\eqref{comrels} imply for the oscillators $a^{\lambda}_{\vec p}$ and $a^{\lambda\dagger}_{\vec p}$ in the mode expansion in Eq.~\eqref{fouriermodes}, note that the latter can be 
inverted using the orthogonality relations \eqref{orthog1} and \eqref{orthog2} as \cite{don-pat-potting2}
\begin{subequations}
\begin{align}
g^{\lambda\lambda'}a^{\lambda'}_{\vec q}&=
i\int d^3x e^{iq\cdot x}\Bigl[
{\stackrel{\leftrightarrow}{\partial_0}}
\left(\eta^{\mu\nu}-(1-\xi^{-1})\delta^\mu_0\delta^\nu_0\right)\nonumber\\
&\qquad\qquad{}-(1-\xi^{-1})q^j(\delta^\mu_j\delta^\nu_0+\delta^\mu_0\delta^\nu_j)
+2k_{AF\,\kappa}\epsilon^{\kappa0\mu\nu}
\Bigr]e_\nu^{*\,(\lambda)}(\vec q) A_\mu(x)\,,\\
g^{\lambda\lambda'}a^{\lambda'\dagger}_{\vec q}&=
-i\int d^3x e^{-iq\cdot x}\Bigl[
{\stackrel{\leftrightarrow}{\partial_0}}
\left(\eta^{\mu\nu}-(1-\xi^{-1})\delta^\mu_0\delta^\nu_0\right)\nonumber\\
&\qquad\qquad{}-(1-\xi^{-1})q^j(\delta^\mu_j\delta^\nu_0+\delta^\mu_0\delta^\nu_j)
+2k_{AF\,\kappa}\epsilon^{\kappa0\mu\nu}
\Bigr]e_\nu^{(\lambda)}(\vec q) A_\mu(x)\, ,
\end{align}
\label{invertedoscillat}%
\end{subequations}
where $q^0 = E_\lambda(\vec{q})$ in both expressions.
Using Eq.~\eqref{invertedoscillat}, together with the
commutation relations in Eqs.~\eqref{comrels}, it can be shown that
the oscillators satisfy the commutation relations
\begin{subequations}
\begin{align}
[a^{\lambda}_{\vec p}, a^{\lambda^\prime\dagger}_{\vec q}] &=
- (2\pi)^3 g^{\lambda \lambda^\prime}\left.\Lambda'_\lambda(p)\delta^3(\vec p - \vec q)\right|_{p^0 = E_\lambda(\vec{p})}\,, \label{oscillatorcom1}\\
[a^{\lambda}_{\vec p}, a^{\lambda^\prime}_{\vec q}]
&=[a^{\lambda\dagger}_{\vec p}, a^{\lambda^\prime\dagger}_{\vec{q}}] = 0\, .\label{oscillatorcom2}
\end{align}
\label{oscillatorcom}%
\end{subequations}
The normalization of the polarization vectors in Eqs.~\eqref{normalizations}, together with the normalization factor $1/\Lambda'_\lambda(p)$ in the definition of the photon field, makes sure that the right-hand side of Eq.~\eqref{oscillatorcom1} is always positive if $\lambda = \lambda' = 3,+,-$, while it is negative if $\lambda = \lambda' = 0$. This holds in all observer frames, and follows from the fact, discussed in Section~\ref{sec:energypositivity}, that $\left.\Lambda'_\lambda(p)\right|_{p^0 = E_\lambda(\vec{p})}$ is always positive.

We define the one-particle state by
\begin{equation}
|\vec{p},\lambda \rangle = a^{\lambda \dagger}_{\vec p} | 0 \rangle\ ,
\label{oneparticlestate}%
\end{equation}
where $|0\rangle$ is the vacuum state that is annihilated by $a^\lambda_{\vec p}$. As in the usual case, the one-particle states with $\lambda = 0$ have negative norm, while the other polarizations have a positive norm. This holds in any observer frame, due to the normalization in Eq.~\eqref{oscillatorcom} and the on-shell form of $g^{\lambda\lambda'}$, given in Eq.~\eqref{polarmetric}. The consistency of the quantization in arbitrary observer frames thus crucially depends on the fact that the sign of $\Lambda_\lambda'(p)$ is an observer Lorentz invariant quantity. One might think that a different choice for the normalization of the polarization vectors or the photon field could invalidate this statement. However, to keep covariant transformation properties for the photon field these changes have to be related and a different choice leads to the same conclusion.

As in the conventional Gupta-Bleuler method one can now go on and implement a gauge-fixing condition on the Hilbert space of physical states, such that no negative-norm states appear in physical observables. In Section~\ref{brst} we show, in the context of BRST quantization, that this can be done consistently.

Finally we note that, although the theory contains negative-energy states in some observer frames, the vacuum is stable in the sense that it is not possible to create physical particles from nothing. This follows from the fact that frames exist in which the theory does not contain any negative-energy states (concordant frames). In such frames energy conservation prohibits the mentioned process. Observer Lorentz invariance then implies that it must be forbidden in any observer frame. A similar argument shows that a charged particle emitting Cherenkov radiation will stop doing so after a while. In concordant frames this happens when it has lost energy to the point that no more photons (with spacelike momenta) can be emitted \cite{kostelecky-lehnert}. This depends on the fact that the energy is bounded from below in all observer frames.

\section{Causality}\label{causality}

The notion of causality is closely related to relativity and Lorentz symmetry. In quantum field theory one usually considers microcausality, i.e. the local (anti)commutativity of observables for spacelike separations. In the present case the theory is microcausal if
\begin{equation}
D^{\mu\nu}(x-y) = \left[A^\mu(x),A^\nu(y)\right] = 0 \qquad {\rm for}\quad (x-y)^2 < 0\ .
\label{microcausality}%
\end{equation}
In this section we will confirm by explicit calculation that Eq.~\eqref{microcausality} holds.

Using the commutation relation of the creation and annihilation operators in Eqs.~\eqref{oscillatorcom}, we find that we can write
\begin{equation}
D^{\mu\nu}(z) = -\int\frac{d^3p}{(2\pi)^3}\sum_{\lambda}\left.\frac{g^{\lambda\lambda}}{\Lambda'_\lambda(p)}\left[e^{(\lambda)\mu} (\vec{p}) e^{(\lambda)\nu *}(\vec{p})\; e^{-ip\cdot z} - e^{(\lambda)\mu *}(\vec{p}) e^{(\lambda)\nu}(\vec{p})\; e^{ip\cdot z}\right]\right|_{p^0 = E_\lambda(\vec{p})}\ ,
\end{equation}
where $z = x-y$. Using Eq.~\eqref{selectionfunction}, the fact that $e^{(\lambda)\mu *}(-p)e^{(\lambda)\nu}(-p) = e^{(\lambda)\mu}(p)e^{(\lambda)\nu *}(p)$, and $\Lambda'(p) \stackrel{p\rightarrow - p}{\longrightarrow} -\Lambda'(p)$, we can write this as
\begin{equation}
D^{\mu\nu}(z) = -\int\frac{d^4p}{(2\pi)^3}\sum_{\lambda\lambda}g^{\lambda\lambda}e^{(\lambda)\mu} (p) e^{(\lambda)\nu *}(p)\; {\rm sgn}(\Lambda'(p)) \delta(\Lambda_\lambda(p))e^{-ip\cdot z}\ ,
\label{causal1aaa}
\end{equation}
It is straightforward to check, by explicit calculation or by using the relation in Eq.~\eqref{PolarProp}, that this is equal to
\begin{equation}
D^{\mu\nu}(z) = -\int\frac{d^3p}{(2\pi)^3}\int_{\mathcal{C}}\frac{dp^0}{(2\pi)}P^{\mu\nu} e^{-ip\cdot z} = i\int\frac{d^3p}{(2\pi)^3}\int_{\mathcal{C}}\frac{dp^0}{(2\pi)}\frac{{\rm Adj(S)^{\mu\nu}}}{\det(S)} e^{-ip\cdot z} ,
\label{causal1aa}
\end{equation}
where the contour in the complex $p^0$ plane encircles all poles in the clockwise direction, $P^{\mu\nu}$ is given in Eq.~\eqref{propagatormomentumspace}, and $\Adj(S)^{\mu\nu} = \det(S)(S^{-1})^{\mu\nu}$ is the adjugate matrix of $S$, whose relation to $P^{\mu\nu}$ is given in Eq.~\eqref{deltatoS}. Note that $P^{\mu\nu}$ contains double and triple poles at $p^\mu = \varsigma \mathcal{K}^\mu$, with $\mathcal{K}^\mu$ defined in Eq.~\eqref{specialmomentum} and $\varsigma = \pm 1$. The result of calculating the residues at these higher-order poles corresponds to the definitions of the polarization vectors at the mentioned momenta, given in Eq.~\eqref{singvectors}.

The expression in Eq.~\eqref{causal1aa} is manifestly observer Lorentz covariant. Therefore, we can calculate its components in a particular frame. If $z^2 = (x-y)^2 < 0$, we can go to an observer frame, where $z^0 = 0$. In this frame we perform the contour integration. Realizing that $\det(S) = \frac{1}{\xi}\prod_{i=0}^8(p^0-\omega_i)$, we get that
\begin{eqnarray}
\left. D^{\mu\nu}(z)\right|_{z^0 =0} &=& -\xi\int \frac{d^3p}{(2\pi)^3} \sum_{i=1}^8 \left[\frac{\Adj(S)^{\mu\nu}}{\prod_{j\neq i}(p^0 - \omega_j)} e^{-i\vec{p}\cdot \vec{z}}\right]_{p^0=\omega_i}\ .
\label{causal1b}
\end{eqnarray}
This result is not valid at the two points in momentum space $p^\mu = \varsigma \mathcal{K}^\mu$ (see Eq.~\eqref{specialmomentum}) when $k_{AF}^2 >0$. At these two momenta, the functions $\Lambda_\lambda(p)$ with $\lambda = 3,+,-$ have degenerate roots. Therefore, the expression in Eq.~\eqref{causal1aa} has a triple pole. However, performing the $p^0$ contour integration at the fixed three-momentum value
$\vec p = \varsigma \vec{\mathcal{K}}$ in the appropriate way gives identically zero. 

Away from $p^\mu = \varsigma \mathcal{K}^\mu$, we use that every component of the numerator in Eq.~\eqref{causal1b} is a polynomial in $p^0$. Furthermore, it is easy to show that
\begin{equation}
\sum_{i=1}^8\frac{(\omega_i)^n}{\prod_{j\neq i}(\omega_i-\omega_j)} = \left\{\begin{array}{cl}0 & {\rm if}\ n=0,\ldots, 6 \\ 1 &{\rm if}\ n=7 \\ \sum_{i=1}^8 \omega_i &{\rm if}\ n=8\end{array}\right.\ .
\end{equation}
Using the explicit expression for $\Adj(S)$, that follows from Eq.~\eqref{deltatoS} and Eq.~\eqref{propagatormomentumspace}, it becomes clear that
\begin{equation}
\left. D^{\mu\nu}(z)\right|_{z^0 =0} = 0\ ,
\end{equation}
i.e. every compenent of $D^{\mu\nu}(z)$ vanishes in an observer frame where $z^0 = 0$. Therefore, since $D^{\mu\nu}(z)$ is observer Lorentz covariant, we conclude that it vanishes in any frame with $(x-y)^2 < 0$, i.e. the fields commute for spacelike separation, confirming microcausality. Notice that we also confirmed Eq.~\eqref{comrels1}. The other commutation relations in Eq.~\eqref{comrels} can be derived in a completely analogous way.

\section{Feynman propagator}\label{feynprop}%

We take the Feynman propagator in coordinate space to be equal to the vacuum expectation value of the time-ordered product of fields at $x$ and $y$, i.e.
\begin{equation}
D^{\mu\nu}_F(x-y) = \theta(x^0 - y^0) D^{\mu\nu}_+(x-y) + \theta(y^0 - x^0) D^{\mu\nu}_-(x-y)\ ,
\end{equation}
with
\begin{subequations}
\begin{eqnarray}
D^{\mu\nu}_+(x-y) &=& \left\langle 0 | A^\mu(x) A^\nu(y) | 0 \right\rangle\ , \\
D^{\mu\nu}_-(x-y) &=& \left\langle 0 | A^\nu(y) A^\mu(x) | 0 \right\rangle\ .
\end{eqnarray}
\end{subequations}
Notice that the $\pm$ signs in these definitions have nothing to do with the $\lambda = \pm$ polarizations, rather they correspond (in concordant frames) to the positive and negative energy modes. This is further clarified when we look at the explicit expressions for $D^{\mu\nu}_\pm(x-y)$ that follow from inserting Eq.~\eqref{fouriermodes}. They are given by
\begin{equation}
D^{\mu\nu}_\pm(z) = -\int\frac{d^4 p}{(2\pi)^4}\sum_{\lambda\lambda'}g^{\lambda\lambda'}\theta(\pm\Lambda'_\lambda(p)){\rm sgn}(\pm\Lambda'_\lambda(p))(2\pi)\delta(\Lambda_\lambda(p))\epsilon^{(\lambda)\mu}\epsilon^{(\lambda')\nu*}e^{-ip\cdot z}\ ,
\end{equation}
with $z = x-y$. Due to the Heaviside stepfunction $\theta(\pm\Lambda'_\lambda(p))$, discussed at the end of Section~\ref{sec:disprel}, $D_\pm^{\mu\nu}(z)$ is only non vanishing if $\pm\Lambda'_\lambda(p) > 0$, which in concordant frames is equivalent to $\pm p^0 > 0$. 

Using the Fourier transform of the Heaviside stepfunction, $\theta(z^0) = \frac{i}{2\pi}\int \frac{e^{-i\tau z^0}d\tau}{\tau + i\varepsilon}$, we can write the Feynman propagator as
\begin{eqnarray}
D^{\mu\nu}_F(z) &=& \int_C \frac{d^4 p}{(2\pi)^4}P^{\mu\nu}e^{-ip\cdot z} \notag \\
&=& -i\int \frac{d^4 p}{(2\pi)^4}\Bigg[\frac{(p^2-m_\gamma^2)\eta^{\mu\nu}}{\Lambda_T(p)-i\varepsilon} + \frac{4\left(p^\mu p^\nu k_{AF}^2 + k_{AF}^\mu k_{AF}^\nu p^2 - (p^\mu k_{AF}^\nu + p^\nu k_{AF}^\mu)(p\cdot k_{AF}) \right)}{(p^2-m_\gamma^2+ i\varepsilon)(\Lambda_T(p)-i\varepsilon)} \notag \\
&& + \frac{2i\epsilon^{\mu\nu\alpha\beta}(k_{AF})_\alpha p_\beta}{\Lambda_T(p)-i\varepsilon} -(1-\xi)\frac{p^\mu p^\nu}{(p^2-\xi m_\gamma^2 + i\varepsilon)(p^2- m_\gamma^2 + i\varepsilon)}\Bigg] e^{-ip\cdot z}  \ ,
\label{Feynmanprop}
\end{eqnarray}
where, on the first line, the integration contour in the complex $p^0$ plane goes above (below) the poles ($\omega$) for which $\left.\Lambda'_\lambda(p)\right|_{p^0 = \omega}$ is positive (negative). After the second equality sign, this is represented by a Feynman $\varepsilon$ prescription.

\section{BRST and the space of states}
\label{brst}

The structure of the space of states is most clearly established in the
BRST formalism \cite{BRST} by
completing the photon Lagrangian (\ref{lagrangian-A})
with the contributions for the St\"uckelberg scalar field
$\phi$ as well as the (anticommuting)
ghost and antighost fields $c$ and $\bar c$
(with ghostnumbers $1$ and $-1$, respectively):
\begin{equation}
\mathcal{L}_{\rm St\ddot{u}ck}=\mathcal{L}_A + \mathcal{L}_\phi + \mathcal{L}_{gh}
\label{lagrangian-Stueck}
\end{equation}
where $\mathcal{L}_A$ is given by (\ref{lagrangian-A}),
while
\begin{equation}
\mathcal{L}_\phi=\tfrac12(\partial_\mu\phi)^2-\tfrac12 \xi m_\gamma^2\phi^2
\end{equation}
and
\begin{equation}
\mathcal{L}_{gh}=-\bar c(\partial^2+\xi m_\gamma^2)c\,.
\end{equation}
Note that the antighost field $\bar c$ is defined to be anti-hermitian
($\bar c^\dagger = -\bar c$), while all other fields are hermitian.
Lagrangian (\ref{lagrangian-Stueck}) can now be obtained from
the Lagrangian
\begin{align}
\mathcal{L'}_{\rm St\ddot{u}ck}&=-\frac14 F_{\mu\nu} F^{\mu\nu}
+ \frac12 k_{AF}^{\kappa} \epsilon_{\kappa\lambda{\mu\nu}}A^{\lambda}F^{{\mu\nu}}  
+\frac12 m_\gamma^2 (A_\mu- \frac{1}{m_\gamma}\partial_\mu\phi)^2 \nonumber\\
&\qquad\quad{}+\frac{\xi}{2}B^2+B(\partial_\mu A^\mu+\xi m_\gamma\phi)
-\bar c(\partial^2+\xi m_\gamma^2)c
\label{lagrangian-Stueck-2}
\end{align}
upon integrating out the (auxiliary) Nakanishi-Lautrup field B \cite{NakLau}.

Lagrangian (\ref{lagrangian-Stueck-2}) changes by a total derivative under the
BRST transformation $s$ defined by
\begin{align}
s A_\mu&=\epsilon\partial_\mu c \label{sA}\\
s \phi&=\epsilon m_\gamma c \label{schi}\\
s \bar c&=\epsilon B \label{scbar} \\
s B&=sc=0
\label{sB}
\end{align}
where $\epsilon$ is some constant infinitesimal Grassmann-valued parameter.
The BRST transformations (\ref{sA})--(\ref{sB}) are generated
by the action of the nilpotent BRST charge
$Q_B = \int d^3 \vec x j^B_0= \int d^3x (B\,\partial_0c-\partial_0B\,c)$,
where $j^B_\mu$ is the conserved Noether current. 
The space of physical states is defined by the space of
closed states (those that are annihilated by $Q_B$) of ghost number zero
modulo the exact states (those that are in the image of $Q_B$).
Restricting ourselves to ghost number zero (no ghost excitations),
it follows from (\ref{schi}) that one-particle states created
by the field $\phi$ are unphysical.
Moreover, we see from (\ref{sA}) and (\ref{schi}) that the linear combination
\begin{equation}
\chi_\mu = A_\mu-\frac{1}{m_\gamma}\partial_\mu \phi
\end{equation}
(the Proca field) is BRST invariant,
and thus any one-particle states created by $\chi_\mu$ are closed.
Finally, we see from (\ref{scbar}) that any one-particle excitations of the
Nakanishi-Lautrup field $B$ are exact.
Using the equations of motion for $B$ and $\phi$,
it follows that on-shell we can replace $B \to \partial_\mu\chi^\mu$.
From this we see that the physical one-particle states can be taken
to correspond to the three transverse polarizations of $\chi_\mu$
(which coincide with the transverse polarizations
$e^{(i)}_\mu(\vec p)$, $i=+,-,3$, of $A_\mu$).
The exact one-particle states correspond to the remaining
longitudinal mode of $\chi_\mu$.

It is worthwhile to point out that the quantization is unaffected
by the Lorentz-violating $k_{AF}$ term.

\section{Discussion}\label{sec:conclusions}%

In this paper, we performed the covariant quantization of Lorentz- and CPT-violating Maxwell-Chern-Simons theory for spacelike, lightlike, as well as timelike $k_{AF}^\mu$. To avoid imaginary energies and for regularization purposes, a non-zero photon mass was introduced through the St\"uckelberg mechanism. This can be done well below any observational constraints. 

We found explicit expressions for a set of four orthogonal and normalized polarization vectors, whose definition is valid in almost all of four-momentum space. These polarization vectors are eigenvectors of the equation-of-motion operator and have the functions $\Lambda_\lambda(p)$, defined in Eq.~\eqref{eigenvalues}, as their eigenvalues. The relations $\Lambda_\lambda(p) = 0$ determine the dispersion relations for the different polarization modes and their solutions fix the on-shell polarization vectors.  The hypersurface in momentum space where the definitions of the polarization vectors are invalid only intersects the relevant mass shells at two singular points and only for timelike $k^\mu_{AF}$. This corresponds to the vanishing of the LV term in the original Lagrangian. We highlighted the treatment of these singular points throughout the paper.

We discussed several properties of the dispersion relation. In particular, we showed that it has eight nondegenerate roots, except at the two singular points, where it has two sets of three degenerate roots (and two nondegenerate ones). We confirmed that the observer-Lorentz-invariant condition $k_{AF}^2 < m_\gamma^2$ guarantees the reality of all the roots. Moreover, we derived an observer Lorentz non-invariant condition, $(k_{AF}^0)^2 < m_\gamma^2$, that makes sure that all energies are positive. Since the latter condition cannot be maintained in arbitrary observer frames, the sign of the roots cannot be used to select a branch of the dispersion relation, as is done in the usual, Lorentz-symmetric, case. However, we found that the sign of $\Lambda'(p)$ can be used instead. This fact is closely related to the observer invariance of the root ordering and the group velocity being smaller than unity.

Being able to unambiguously identify the different branches of the dispersion relation in all observer frames, allowed us to construct an observer Lorentz covariant mode expansion of the photon field in terms of the polarization vectors for the different modes. Using the resulting explicit expression for the photon field, we performed the quantization of the theory. We also derived the Feynman propagator and showed that the theory is microcausal. Finally we showed, in the context of BRST quantization, that the three transverse modes are the physical ones.

One obvious direction into which one can extend the present work is investigating the massless limit. We expect that taking the limit $m_\gamma \rightarrow 0$ at the end of a calculation of a physical observable gives a consistent result for lightlike and spacelike $k_{AF}^\mu$ (for timelike $k_{AF}^\mu$, the imaginary energies will reappear). All the more because the $\lambda = 3$ polarization mode seems to decouple in a gauge-invariant theory, because $e^{(3)\mu} \propto p^\mu$ in that case, resulting in two physical states. As in the Lorentz-symmetric case, in the massless limit it is not possible to find a basis of four orthogonal covariant polarization vectors in the general class of gauges we consider in this paper. However, we expect that it is possible, using BRST quantization, to show that the non-covariant components of the field are unphysical and decouple.

A second option for follow-up work is to include interactions and quantum effects. The latter might introduce other, possibly higher-dimensional, LV coefficients through radiative corrections. To go beyond tree-level one has to consider the effect of such effective LV coefficients.

Finally, one could try to apply the methods of the present work to the CPT-even $k_F$ term of the minimal SME or even include higher-dimensional kinetic terms for the photon, which have been categorized in Ref.~\cite{higherdim}.
Note that, in the latter case, 
one would have to find a way to consistently deal with spurious Ostrogradski modes \cite{Ostro} that arise due to higher-order time derivatives in the Lagrangian.

Presently, the fact that the covariant quantization of the present Lorentz- and CPT-violating theory is possible, at least with a non zero photon mass (well below observational constraints), despite the presence of negative-energy states in some observer frames, is an important result of this paper. It is of relevance, in particular, to considerations of vacuum Cherenkov radiation, for which such negative-energy states are unavoidable. Moreover, the explicit expressions for the polarization vectors, their bilinears, and the Feynman propagator, in arbitrary observer frames, pave the way for calculations of LV observables involving $k_{AF}^\mu$.

\acknowledgments
We thank V.~A.~Kosteleck\'y for helpful suggestions.
This work is supported in part by the Funda\c c\~ao para a Ci\^encia e a Tecnologia of Portugal (FCT) through projects UID/FIS/00099/2013 and SFRH/BPD/101403/2014 and program POPH/FSE.

\appendix

\section{Reality and sign of the roots of $\Lambda_\pm(p)$} \label{appA}

In this Appendix we give the details of some statements made in the main text, concerning the sign and the possible complex-valuedness of the roots of $\Lambda_\pm(p)$. We do this by considering the functions defined in Eq.~\eqref{partfunctions} and plotted in Figs.~\ref{fig:timelikek} and \ref{fig:spacelikek} for the case of timelike and spacelike/lightlike $k^\mu_{AF}$, respectively.

First of all, we discuss the condition
\begin{equation}
|x_\alpha| < \sqrt{\vec{p}^2 + m_\gamma^2}
\label{appcond1}
\end{equation}
for spacelike and lightlike $k^\mu_{AF}$, which is used below Eq.~\eqref{xplusmin}. The points $p^0 = x_\alpha$ are the points where the branches of $\pm f_{\delta}(p^0)$ start (see Fig.~\ref{fig:spacelikek}) and their expressions are given in Eq.~\eqref{xplusmin}. Eq.~\eqref{appcond1} reflects the condition that these points stay inside the curve of $f_0(p^0)$. A little algebra shows that Eq.~\eqref{appcond1} holds if
\begin{equation}
\vec{p}^2 > \frac{-\vec{k}_{AF}^2 m_\gamma^2}{(k^0_{AF}\sqrt{\sin^2\theta}+\alpha \sqrt{-k_{AF}^2}\cos\theta)^2}\ ,
\label{appcond1b}
\end{equation}
where $\theta$ is the angle between $\vec{p}$ and $\vec{k}_{AF}$. Since the expression on the right is always negative, we see that Eqs.~\eqref{appcond1} and \eqref{appcond1b} are always satisfied and thus that $\Lambda_{\pm}(p)$ always have two real roots each for spacelike and lightlike $k^\mu_{AF}$. 

Next, we consider the inequality
\begin{equation}
-f_\delta(0) > f_0(0)\ ,
\label{appcond2}
\end{equation}
which, for $k_{AF}^2 > 0$, is sufficient to make sure that the signs of $\omega_5$ and $\omega_6$ are different. For $k_{AF}^2 < 0$ it can happen that $f_\delta(0)$ is not real. However, in that case it is obvious from Fig.~\ref{fig:spacelikek} and the argument following Eq.~\eqref{appcond1} that the signs of $\omega_5$ and $\omega_6$ will differ. It is easy to see that $-f_\delta(0) = f_0(0)$ if
\begin{equation}
|\vec{p}| = \sqrt{(k_{AF}^0)^2 - \vec{k}_{AF}^2\sin^2\theta} \pm \sqrt{(k_{AF}^0)^2 - \vec{k}_{AF}^2\sin^2\theta - m_\gamma^2}\ .
\label{appcond3}
\end{equation}
If $k^\mu_{AF}$ is spacelike, the first square root can become imaginary. This corresponds to $f_\delta(0)$ being imaginary, for which case $\omega_5$ and $\omega_6$ differ in sign. If both square roots in Eq.~\eqref{appcond3} are real, Eq.~\eqref{appcond3} defines an interval for $|\vec{p}|$, outside of which Eq.~\eqref{appcond2} is satisfied and $\Lambda_+(p)$ has two roots of opposite sign. Inside of the interval, however, the signs of $\omega_5$ and $\omega_6$ are the same and after the redefinition of one of the roots, the theory can contain states of negative energy. It is clear from Eq.~\eqref{appcond3} that there will be no negative energies if $(k_{AF}^0)^2 < m_\gamma^2$, because the second square root is imaginary in that case. This confirms that the condition in Eq.~\eqref{normalsol} implies energy positivity. Furthermore, the second square root is real if 
\begin{equation}
\cos^2\theta > \frac{m_\gamma^2 - k_{AF}^2}{\vec{k}_{AF}^2}\ .
\label{appcond3b}
\end{equation}
For angles satisfying this inequality and $|\vec{p}|$ in the interval in Eq.~\eqref{appcond3}, the two roots of the dispersion relation $\Lambda_+(p) = 0$ have the same sign. Since we redefine the energies as in Eq.~\eqref{energies}, the theory will contain negative-energy photons if $\omega_5(\vec{p}) < 0$. This happens if the extremum of $f_\delta(p^0)$ lies to the left of $p^0 = 0$, i.e. if $k^0(\vec{p}\cdot \vec{k}_{AF}) < 0$. The momentum of the photons with negative energy thus lies in a cone around the direction defined by $-{\rm sgn}(k^0)\vec{k}_{AF}$ (and not in the opposite direction).

Finally, we show that all roots of $\Lambda_+(p)$ are real if Eq.~\eqref{normalsol2} holds, i.e. if
\begin{equation}
k_{AF}^2 < m_\gamma^2\ .
\end{equation}
For lightlike and spacelike $k_{AF}^\mu$, this was already evident from the considerations following Eq.~\eqref{appcond1}. In the following, we show it for timelike $k_{AF}^\mu$.
To achieve this, we ascertain that if  $k_{AF}^2 < m_\gamma^2$, we can always find a value of $p^0$ for which
\begin{equation}
-f_\delta(p^0) > f_0(p^0)\ ,
\end{equation}
meaning that $-f_\delta(p^0)$ must intersect $f_0(p^0)$ at two different values of $p^0$, corresponding to the two real roots of $\Lambda_+(p)$. To proof this, we start with an ansatz for $p^0$:
\begin{equation}
p^0 = a|\vec{p}|\ ,
\end{equation}
with $a$ a dimensionless factor. At this value of $p^0$, we find
\begin{subequations}
\begin{eqnarray}
f_0(a|\vec{p}|) &=& \vec{p}^2(a^2 - 1) - m_\gamma^2\ , \\
-f_\delta(a|\vec{p}|) &=& -2|\vec{p}|\sqrt{X} \equiv -2|\vec{p}|\sqrt{a^2\vec{k}_{AF}^2 - 2ak_{AF}^0|\vec{k}_{AF}|\cos\theta + \vec{k}_{AF}^2\cos^2\theta + k_{AF}^2}\ .
\end{eqnarray}
\end{subequations}
We do not gain much insight by solving $-f_\delta(a|\vec{p}|) > f_0(a|\vec{p}|)$ for $a$ directly. However, we easily find that $-f_\delta(a|\vec{p}|) = f_0(a|\vec{p}|)$ for 
\begin{equation}
|\vec{p}| = \frac{\sqrt{X}\pm \sqrt{X + (a^2 - 1)m_\gamma^2}}{1-a^2}\ .
\label{appcond4}
\end{equation}
Furthermore, $-f_\delta(a|\vec{p}|) - f_0(a|\vec{p}|)$, as a function of $|\vec{p}|$, is a parabola that opens upward if $a^2 < 1$. So, if $a^2 < 1$ and $|\vec{p}|$ is in the interval defined by Eq.~\eqref{appcond4}, then $-f_\delta(a|\vec{p}|) < f_0(a|\vec{p}|)$. It follows that if we can find an $|a| < 1$ such that the second square root in Eq.~\eqref{appcond4} becomes imaginary, then the mentioned interval of $|\vec{p}|$ does not exist and therefore $-f_\delta(a|\vec{p}|) > f_0(a|\vec{p}|)$ for any $|\vec{p}|$.

We find that the argument of the second square root in Eq.~\eqref{appcond4} vanishes if
\begin{equation}
a = \frac{(k_{AF}^0)^2}{m_\gamma^2+\vec{k}_{AF}^2}\left(\frac{|\vec{k}_{AF}|}{k_{AF}^0}\cos\theta \pm \frac{1}{k_{AF}^0}\sqrt{(m_\gamma^2 - k_{AF}^2)\left(\frac{m_\gamma^2 + \vec{k}_{AF}^2}{(k^0_{AF})^2} - \cos^2\theta\frac{\vec{k}_{AF}^2}{(k^0_{AF})^2} \right)}\right)\ .
\label{appcond5}
\end{equation}
It is straightforward to check that the square root is real if $k_{AF}^2<m_\gamma^2$ and that the absolute value of first term is smaller than one in that case. Eq.~\eqref{appcond5} thus defines an $a$-interval for which the second square root in Eq.~\eqref{appcond4} is imaginary. Therefore, near the center of this interval, there are values of $|a|<1$ for which $-f_\delta(a|\vec{p}|) > f_0(a|\vec{p}|)$. This means that $-f_\delta(p^0)$ intersects $f_0(p^0)$ at two different values of $p^0$. We conclude that $\Lambda_+(p)$ always has two real roots if $k_{AF}^2 < m_\gamma^2$. 

\section{Group velocity} \label{appB}

In this appendix, we consider the group velocity of the different modes of the photon. It is defined in Eq.~\eqref{groupvelocity} as
\begin{equation}
\vec{v}_g^{(\lambda)} = \frac{\partial E_\lambda(\vec{p})}{\partial \vec{p}}\ .
\label{groupvelocity3}
\end{equation}
We will show that 
\begin{equation}
\vec{v}_g^{(\lambda)} = -\left[\frac{\partial \Lambda_\lambda(p)}{\partial \vec{p}}\bigg/ \frac{\partial \Lambda_\lambda(p)}{\partial p^0}\right]_{p^0 = E_\lambda(\vec{p})}
\label{groupvelocity2}
\end{equation}
and that $|\vec{v}_g^{(\lambda)}| < 1$. For the modes with $\lambda = 0,3$ this is trivial and we will only consider the $\lambda = \pm$ modes in the remainder of this appendix. 

The fact that Eq.~\eqref{groupvelocity2} holds, follows easily by realizing that $\Lambda_T(p) = \Lambda_+(p)\Lambda_-(p)$ is a polynomial in $p^0$, which allows us to write it as 
\begin{equation}
\Lambda_T(p) = (p^0 - E_+(\vec{p}))(p^0 + E_+(-\vec{p}))(p^0 - E_-(\vec{p}))(p^0 + E_-(-\vec{p}))\ ,
\end{equation}
where we used the energy redefinitions, given in Eq.~\eqref{energies}. From Eq.~\eqref{groupvelocity2}, it follows that, for $\lambda = +,-$,

\begin{equation}
\frac{\partial E_\pm(\vec{p})}{\partial \vec{p}} = -\left[\frac{\partial \Lambda_T(p)}{\partial \vec{p}}\bigg/ \frac{\partial \Lambda_T(p)}{\partial p^0}\right]_{p^0 = E_\pm(\vec{p})} = -\left[\frac{\partial \Lambda_\pm(p)}{\partial \vec{p}}\bigg/ \frac{\partial \Lambda_\pm(p)}{\partial p^0}\right]_{p^0 = E_\pm(\vec{p})}\ ,
\end{equation} 
confirming Eq.~\eqref{groupvelocity2}. This equality does not hold if $p^\mu = \varsigma\mathcal{K}^\mu$. In that degenerate case, the group velocity, as given in Eq.~\eqref{groupvelocity3}, becomes ill-defined, as can be seen from explicit calculations for purely timelike $k^\mu_{AF}$, or from the analysis at the end of this section. However, we can assign a value to the right-hand side of Eq.~\eqref{groupvelocity2} by employing some limiting procedure, as described below Eqs.~\eqref{singvectors}. This is in fact the quantity we need in Eq.~\eqref{nosignchange}.

It remains to be shown that $|\vec{v}_g^{(\lambda)}| < 1$. To this affect we define
\begin{equation}
w_\pm^\mu \equiv \frac{\partial \Lambda_\pm(p)}{\partial p_\mu} = 2\left(p^\mu \pm \frac{(p\cdot k_{AF})k_{AF}^\mu - k_{AF}^2 p^\mu}{\sqrt{(p\cdot k_{AF})^2 - p^2 k_{AF}^2}}\right)\ ,
\label{wdef}
\end{equation}
such that $\vec{v}_g^{(\lambda)} = -\left[\vec{w}_\lambda / w_\lambda^0\right]_{p^0 = E_\lambda(\vec{p})}$. It follows that $|\vec{v}_g^{(\lambda)}| < 1$ holds, if $w_\lambda^\mu$, evaluated on shell, is timelike. We determine that on shell $w_\lambda^2$ is given by
\begin{equation}
\left. w_\lambda^2\right|_{p^0 = E_\lambda(\vec{p})} =4(m_\gamma^2 - k_{AF}^2)\ .
\label{wtimelike}
\end{equation}
Therefore, $w_\lambda^2 > 0$ if $k_{AF}^2 < m_\gamma^2$. The latter is a necessary condition for the theory to be consistent, as already mentioned in Section~\ref{sec:disprel}. We thus conclude that the absolute value of photon group velocity is smaller than unity for the cases we consider. For the degenerate case of $p^\mu = \varsigma\mathcal{K}^\mu$, this statement is invalid, because Eq.~\eqref{groupvelocity2} does not hold (the group velocity becomes ill-defined). However, Eq.~\eqref{wtimelike} shows that the quantity relevant for Eq.~\eqref{nosignchange}, which is the right-hand side of Eq.~\eqref{groupvelocity2}, is smaller than unity, even if $p^\mu = \varsigma\mathcal{K}^\mu$. We note that a method to desingularize the classical group velocity at the singular points exists \cite{desingularization}.

\section{Energy lower bound} \label{appC}

Using the expression for the group velocity implied by Eqs.~\eqref{groupvelocity2} and \eqref{wdef}, we will determine the lowest value the photon energy can reach in a particular observer frame. For the polarization modes with $\lambda = 0,3$ this is trivial, so we will focus on the $\lambda = \pm$ modes. To find the stationary points of the energy as a function of $\vec{p}$, we will determine the $\vec{p}$ values for which $\vec{w}_\pm$ in Eq.~\eqref{wdef}, vanishes. These points correspond to the lower bound for the energy, unless the energy at the singular point (given by $m_\gamma |k^0_{AF}|/\sqrt{k_{AF}^2}$) is smaller. 

It is straightforward to establish that $\vec{w}_\pm$ vanishes if $\vec{k}_{AF} = \vec{0}$, if $k^0_{AF} = 0$, or if $\vec{p} \propto \vec{k}_{AF}$, i.e. when $(\vec{p}\cdot\hat{k}_{AF})$ is either $|\vec{p}|$ or $-|\vec{p}|$. In all of these cases, the dispersion relation can be solved exactly. For purely timelike and purely spacelike $k^\mu_{AF}$, the concordant-frame-positive energy solutions are given by
\begin{subequations}
\begin{eqnarray}
\left.E_\pm(\vec{p})\right|_{\vec{k}_{AF} = \vec{0}} &=& \sqrt{\vec{p}^2 + m_\gamma^2 \mp 2|k^0_{AF}||\vec{p}|}\ , \\
\left.E_\pm(\vec{p})\right|_{k^0_{AF} = 0} &=& \sqrt{\vec{p}^2 + m_\gamma^2 + 2\vec{k}_{AF}^2 \mp 2\sqrt{\vec{k}_{AF}^4 + m_\gamma^2\vec{k}_{AF}^2 + (\vec{p}\cdot\vec{k}_{AF})^2}}\ ,
\end{eqnarray}
\end{subequations}
such that
\begin{subequations}
\begin{eqnarray}
\left.\vec{v}_g^{(\pm)}\right|_{\vec{k}_{AF} = \vec{0}} &=& \frac{\vec{p} \mp |k^0_{AF}|\hat{p}}{E_\pm(\vec{p})}\ , \label{grouppuretime}\\
\left.\vec{v}_g^{(\pm)}\right|_{k^0_{AF} = 0} &=& \frac{\vec{p} \sqrt{\vec{k}_{AF}^4 + m_\gamma^2\vec{k}_{AF}^2 + (\vec{p}\cdot\vec{k}_{AF})^2} \mp (\vec{p}\cdot\vec{k}_{AF})\vec{k}_{AF}}{E_\pm(\vec{p})\sqrt{\vec{k}_{AF}^4 + m_\gamma^2\vec{k}_{AF}^2 + (\vec{p}\cdot\vec{k}_{AF})^2}}\ . \label{grouppurespace}%
\end{eqnarray}
\end{subequations}
The group velocity for the purely timelike case in Eq.~\eqref{grouppuretime} can only vanish for the $\lambda = +$ mode, in which case $|\vec{p}| = |k^0_{AF}|$, giving a energy lower bound of $\sqrt{m_\gamma^2 - (k^0_{AF})^2}$. For the $\lambda = -$ mode, the energy lower bound for the purely timelike case is the energy at the singular point ($\vec{p} = \vec{0}$), at which the group velocity becomes ill-defined. If $k^0_{AF} = 0$, the group velocity vanishes if $\vec{p} = 0$ (we will deal with $\vec{p} \propto \vec{k}_{AF}$ seperately). It follows that the minimal energy is then given by $\sqrt{m_\gamma^2 + \vec{k}_{AF}^2} \mp |\vec{k}_{AF}|$.

Having dealt with the special cases of purely timelike and purely spacelike $k_{AF}^\mu$, we proceed to the general case where the LV four vector has both nonzero time and space components. As mentioned earlier, the group velocity then only vanishes if $\vec{p}$ is (anti)parallel to $\vec{k}_{AF}$. The corresponding expressions for $E_\pm(\vec{p})$ are given by:
\begin{eqnarray}
\left.E_\pm(\vec{p})\right|_{\vec{p} \propto \vec{k}_{AF}} &=& \left\{\begin{array}{ccc}
\sqrt{\vec{p}^2 + m_\gamma^2 + \vec{k}_{AF}^2 \pm 2k^0_{AF} (\vec{p}\cdot \hat{k}_{AF})} \mp |\vec{k}_{AF}| & {\rm if} & \frac{k^0_{AF} (\vec{p}\cdot \hat{k}_{AF})}{|\vec{k}_{AF}|} \leq \sqrt{\vec{p}^2 + m_\gamma^2} \\
\sqrt{\vec{p}^2 + m_\gamma^2 + \vec{k}_{AF}^2 \mp 2k^0_{AF} (\vec{p}\cdot \hat{k}_{AF})} \pm |\vec{k}_{AF}| & {\rm if} & \frac{k^0_{AF} (\vec{p}\cdot \hat{k}_{AF})}{|\vec{k}_{AF}|} \geq \sqrt{\vec{p}^2 + m_\gamma^2}
\end{array}\right.\ . \notag \\
\label{specialenergies}%
\end{eqnarray}
The lower expression only applies for timelike $k_{AF}^\mu$, because $k^0_{AF}(\vec{p}\cdot \hat{k}_{AF}) > |\vec{k}_{AF}|\sqrt{\vec{p}^2+ m_\gamma^2}$ requires $k_{AF}^2 > 0$. For $k^0_{AF}(\vec{p}\cdot \hat{k}_{AF}) = |\vec{k}_{AF}|\sqrt{\vec{p}^2+ m_\gamma^2}$, which also requires timelike $k_{AF}^\mu$ and corresponds to the singular point of Eq.~\eqref{specialmomentum}, Eq.~\eqref{specialenergies} gives $E_\pm(\vec{p}) = \sqrt{\vec{p}^2 + m_\gamma^2}$, showing that the energy is continuous through the singular point. However, the group velocity that follows from Eq.~\eqref{specialenergies} is given by
\begin{eqnarray}
\left.\vec{v}_g^{(\pm)}\right|_{\vec{p} \propto \vec{k}_{AF}} &=& \left\{\begin{array}{ccc}
\displaystyle\frac{\vec{p} \pm {\rm sgn}(\hat{p}\cdot\hat{k}_{AF})k^0_{AF}\hat{p}}{E_\pm(\vec{p}) \pm |\vec{k}_{AF}|} & {\rm if} & \frac{k^0_{AF} (\vec{p}\cdot \hat{k}_{AF})}{|\vec{k}_{AF}|} \leq \sqrt{\vec{p}^2 + m_\gamma^2} \\
\displaystyle\frac{\vec{p} \mp {\rm sgn}(\hat{p}\cdot\hat{k}_{AF})k^0_{AF}\hat{p}}{E_\pm(\vec{p}) \mp |\vec{k}_{AF}|} & {\rm if} & \frac{k^0_{AF} (\vec{p}\cdot \hat{k}_{AF})}{|\vec{k}_{AF}|} \geq\sqrt{\vec{p}^2 + m_\gamma^2}
\end{array}\right.\ , \notag \\
\label{specialgroupvelocities}%
\end{eqnarray}
which is clearly not continuous through the singular point, because approaching from below gives $\vec{v}_g^{(\pm)} = \left(\frac{|\vec{k}_{AF}|}{|k^0_{AF}|} \pm \frac{\sqrt{k_{AF}^2}}{m_\gamma}\right)\hat{p}$, while approaching from above gives $\vec{v}_g^{(\pm)} = \left(\frac{|\vec{k}_{AF}|}{|k^0_{AF}|} \mp \frac{\sqrt{k_{AF}^2}}{m_\gamma}\right)\hat{p}$. Using methods described in Ref.~\cite{desingularization}, one can nevertheless rigorously define the group velocity at the singular points. Here we instead choose to check explicitly if the energy at the singular point is the smallest energy value.

We can thus use Eq.~\eqref{specialgroupvelocities} to determine the lower bound for the energies, unless the minimal energy is reached exactly at the singular point, which we check explicitly. Investigating when the expression for the group velocity vanishes, while simultaneously satisfying the condition on the right, and comparing to the energy at the singular point, we come to the conclusion that the lower bound for the photon energies in the $\lambda = \pm$ modes is given by
\begin{subequations}
\begin{eqnarray}
\left. E_-(\vec{p})\right|_{\rm min} &=& \left\{\begin{array}{ccc} \frac{m_\gamma |k^0_{AF}|}{\sqrt{k^2_{AF}}} & {\rm if } & |\vec{k}_{AF}| \leq \frac{(k^0_{AF})^2}{\sqrt{(k^0_{AF})^2 + m_\gamma^2}} \label{lowerboundEmin}  \\
\sqrt{m_\gamma^2 - k_{AF}^2} + |\vec{k}_{AF}| & {\rm if} & |\vec{k}_{AF}| \geq \frac{(k^0_{AF})^2}{\sqrt{(k^0_{AF})^2 + m_\gamma^2}}
 \end{array}\right.\ , \\
\left. E_+(\vec{p})\right|_{\rm min} &=& \sqrt{m_\gamma^2 - k_{AF}^2} - |\vec{k}_{AF}|\ , \label{lowerboundEplus}%
\end{eqnarray}
\label{lowerboundE}%
\end{subequations}
for timelike, lightlike, as well as spacelike $k_{AF}^\mu$. Eqs.~\eqref{lowerboundE} also capture the results for the purely timelike and purely spacelike case. 

These results show, first of all, that the energy has a finite, albeit observer dependent, lower bound. Furthermore, we confirm the results of appendix~\ref{appA}, that the energy of the $\lambda = -$ mode is always positive, while $ E_+(\vec{p})$ can become negative if $(k^0_{AF})^2 > m_\gamma^2$.

\begin{thebibliography}{99}
\bibitem{relationtob} 
  The dispersion relation for the dimensionful, CPT- and parity-odd SME fermion parameter $b^\mu$, closely resembles the dispersion relation of the $\lambda = \pm$ modes. In fact, after replacing $k^\mu_{AF} \rightarrow b^\mu$ and $m_\gamma^2 \rightarrow m_f^2 + b^2$ ($m_f$ is the fermion mass), the expression $\Lambda_+(p)\Lambda_-(p)=0$ becomes the $b^\mu$ dispersion relation. The results obtained in Section~\ref{sec:disprel} are thus also valid for LV fermions with a $b^\mu$ parameter. Notice in particular that the conditions in Eqs.~\eqref{normalsol} and \eqref{normalsol2}, become $\vec{b}^2 < m_f^2$ and $m_f^2 > 0$, respectively. This confirms the assertion in Ref.~\cite{kostelecky-lehnert} that the energy of such fermions is always real, but not always positive.
\end{thebibliography}
\end{document}